\documentclass[11pt]{article}

\usepackage[a4paper,left=20mm,top=20mm,right=20mm,bottom=25mm]{geometry}

\usepackage{amsmath, amsfonts, amssymb, amsthm}
\usepackage{mathtools}
\usepackage[mathscr]{euscript}
\usepackage{wasysym}

\usepackage{graphicx}
\usepackage{xcolor}
\usepackage{enumitem}
\usepackage{authblk}

\usepackage{color}
\usepackage[colorlinks=true,urlcolor=black,linkcolor=blue,citecolor=blue]{hyperref}
\usepackage[font=footnotesize,width=.85\textwidth,labelfont=bf]{caption}
\usepackage{underscore}  % to fix doi underscores DO NOT USE _ in file names
\usepackage{mathptmx}
\usepackage[mathscr]{euscript}

\usepackage[normalem]{ulem}

\usepackage{natbib}
\usepackage{bibentry}
\setcitestyle{authoryear,open={(},close={)}}
\let\cite\citep
\usepackage[linesnumbered,ruled,vlined]{algorithm2e}

\newtheorem{theorem}{Theorem}
\newtheorem{lemma}{Lemma}%[section]
\newtheorem{definition}{Definition}%[section]
\newtheorem{corollary}{Corollary}%[section]

\newtheorem{proposition}{Proposition}
\newtheorem{fact}{Observation}

\newtheorem*{claim}{Claim}
\newenvironment{claim-proof}%
{\begin{description}
    \item \emph{Proof of Claim:}}{\hfill$\diamond$\end{description}}

\DeclareMathOperator{\cl}{cl}
\DeclareMathOperator{\lca}{lca}
\DeclareMathOperator{\child}{child}
\DeclareMathOperator{\parent}{par}
\DeclareMathOperator{\depth}{depth}
\DeclareMathOperator{\LA}{LA}

\newcommand{\partTH}{\mathfrak{P}} % partition induced by T and H
\newcommand{\fitch}{\digamma} % fitch graph of T and H
\newcommand{\symfitch}{\overline{\fitch}} % symmetrized fitch graph of T and H
\newcommand{\qfitch}{\mathscr{F}} % fitch quotient graph of T and H
\newcommand{\SPEC}{\newmoon}
\newcommand{\HGT}{\triangle}
\newcommand{\DUPL}{\square}
\newcommand{\LOSS}{\perp}

\newcommand{\TP}{\widehat{T}}

\newcommand{\AX}[1]{\upshape{#1}}
\newcommand{\thmref}[1]{\,\upshape{#1}\quad}

\providecommand{\keywords}[1]{\textbf{\textit{Keywords: }} #1}

\title{Orientation of Fitch Graphs and Detection of Horizontal Gene Transfer in 
Gene Trees}

\author[1,2]{David Schaller}
\author[3]{Marc Hellmuth}
\author[1,2,4,5,6,7]{Peter F.\ Stadler}

\affil[1]{Bioinformatics Group, Department of Computer Science \&
  Interdisciplinary Center for Bioinformatics, Leipzig University,
  H{\"a}rtelstra{\ss}e 16–18, D-04107 Leipzig, Germany
  \authorcr \texttt{sdavid@bioinf.uni-leipzig.de} $\cdot$ 
  \texttt{studla@bioinf.uni-leipzig.de}}

\affil[2]{Max Planck Institute for Mathematics in the Sciences,
  Inselstra{\ss}e 22, D-04103 Leipzig, Germany}

\affil[3]{Department of Mathematics, Faculty of Science, Stockholm
  University, SE - 106 91 Stockholm, Sweden
  \authorcr \texttt{mhellmuth@mailbox.org}}

\affil[4]{German Centre for Integrative Biodiversity Research (iDiv)
  Halle-Jena-Leipzig; Competence Center for Scalable Data Services
  and Solutions; and Leipzig Research Center for Civilization Diseases,
  Leipzig University, Germany}

\affil[5]{Inst.\ f.\ Theoretical Chemistry, University of Vienna,
  W{\"a}hringerstra{\ss}e 17, A-1090 Wien, Austria}

\affil[6]{Facultad de Ciencias, Universidad National de Colombia, Sede
  Bogot{\'a}, Colombia}

\affil[7]{Santa Fe Institute, 1399 Hyde Park Rd., Santa Fe, NM 87501, USA}

\date{\ }

\begin{document}
  
\maketitle 

\abstract{
  Horizontal gene transfer events partition a gene tree $T$ and thus, its
  leaf set into subsets of genes whose evolutionary history is described by
  speciation and duplication events alone. Indirect phylogenetic methods
  can be used to infer such partitions $\mathcal{P}$ from sequence
  similarity or evolutionary distances without any \emph{a priory}
  knowledge about the underlying tree $T$.  In this contribution, we assume
  that such a partition $\mathcal{P}$ of a set of genes $X$ is given and
  that, independently, an estimate $T$ of the original gene tree on $X$ has
  been derived.  We then ask to what extent $T$ and the xenology
  information, i.e., $\mathcal{P}$ can be combined to determine the
  horizontal transfer edges in $T$. We show that for each pair of genes $x$
  and $y$ with $x,y$ being in different parts of $\mathcal{P}$, it can be
  decided whether there always exists or never exists a horizontal gene
  transfer in $T$ along the path connecting $y$ and the most recent common
  ancestor of $x$ and $y$.  This problem is equivalent to determining the
  presence or absence of the directed edge $(x,y)$ in so-called Fitch
  graphs; a more fine-grained version of graphs that represent the
  dependencies between the sets in $\mathcal{P}$.  We then consider the
  generalization to insufficiently resolved gene trees and show that
  analogous results can be obtained. We show that the classification of
  $(x,y)$ can be computed in constant time after linear-time preprocessing.
  Using simulated gene family histories, we observe empirically that the
  vast majority of horizontal transfer edges in the gene tree $T$ can be
  recovered unambiguously.
}

\bigskip
\noindent

\sloppy

\section{Introduction}

% -> added comment to the discussion, I think we don't want to open this
% in the intro already:
%\TODO{mh memo: in https://link.springer.com/article/10.1007/s00285-019-01384-x
%  hatten wir schon gezeigt, dass " .. important biological insight is the
%  observation that the phylogenetic signal to infer these gene trees is
%  entirely contained in the non-HGT events"}

Gene family histories (GFHs) play an important role for the understanding
of innovations in evolutionary biology. A GFH describes the changes in the
set of related genes along the evolutionary history of a set of species of
interest. Mathematically, this amounts to the ``reconciliation'', i.e., the
embedding of a gene tree $T$ into a species tree $S$, such that the nodes
of $T$ describing gene duplication and horizontal gene transfer (HGT)
events are mapped into the edges of $S$, see e.g.\
\cite{HW:16book,Setubal:18a,Altenhoff:19} and the references therein.  Each
transfer edge of $T$ must then be mapped in a way such that its endpoints
are mapped to distinct lineages in $S$.  In practice, the accurate
inference in particular of the gene tree $T$ from sequence data, however,
is a difficult problem plagued by biases and technical artifacts associated
with the necessity to extract additive dissimilarities between genes.

An alternative -- combinatorial -- approach avoids the requirement of
additive distances and instead uses only qualitative comparisons of
pairwise distances \cite{Hellmuth:15a}. In particular, two types of
information have turned out to be useful: the \emph{best matches} of a gene
in the genome of a different species \cite{Geiss:19a}, and the fact that
the last common ancestor of a pair of genes is younger than the last common
ancestor of the species in which they reside \cite{Schaller:21f}. Both
types of data yield vertex-colored graphs whose vertices represent the
genes with colors identifying the species in which they reside. The
\emph{best match} graphs (BMG) \cite{Geiss:19a} are directed. In contrast,
the \emph{later-divergence-time} (LDT) graphs are undirected and capture
HGT events. More precisely, LDTs can be used to find a partition
$\mathcal{P}$ of the set of genes such that within each subset the genes
that are \emph{not} separated by a HGT event since their last common
ancestor, while HGT events separate distinct subsets of this partition. The
best match graph, on the other hand, can be used to infer a minor of the
gene tree $T$, i.e., a not necessarily fully resolved version of $T$.
Alternatively, estimated orthology graphs, which are equivalent to cographs
or, more general, to symbolic ultrametrics \cite{Boecker:98,Hellmuth:13a},
can be used to infer a minor of $T$ \cite{Hellmuth:15a}. It is of practical
interest, therefore, to identify the edges of $T$ that correspond to HGT
events \cite{Jones:17,Schaller:21f}. Here we ask to what extent this is
possible based on a given tree $T$ and a given partition $\mathcal{P}$ of
the leaf set of $T$ into HGT-free subsets.

The \emph{Fitch graph} of a GFH also has the genes as vertices and features
a directed edge from gene $x$ to gene $y$ whenever a horizontal transfer
event was present between $y$ and the last common ancestor of $x$ and $y$
\cite{Geiss:18a}. Consequently, $x$ and $y$ belong to different sets of
$\mathcal{P}$. The symmetrized Fitch graph \cite{Hellmuth:18a} thus is the
complete multipartite graph that has the sets of $\mathcal{P}$ as its
maximal independent sets.  It is
of practical interest, therefore, to investigate to what extent the
partition $\mathcal{P}$ already identifies the placement of the HGT events
in the tree $T$. In related work, we recently characterized under which
conditions an orthology graph, i.e., a symbolic ultrametric, and a Fitch
graph are consistent in the sense that they can be explained by a common
vertex- and edge-annotated gene tree $T$ \cite{Hellmuth:21w}.  Using this
information, a GFH, i.e., a species tree $S$ together with a
\emph{time-consistent} reconciliation map between the annotated gene tree
$T$ and $S$, can then be determined in polynomial time, provided it exists
\cite{Hellmuth:17,nojgaard2018time,lafond2020reconstruction}.  From a
practical point of view, however, this result is of limited use since so
far there is no efficient means of inferring the directed Fitch graph
directly from data, while its symmetrized counterpart is accessible from
empirical data \cite{Schaller:21f}. This prompts us to ask to what extent
the compatibility of $T$ and $\mathcal{P}$ determines the orientation of
the edges in the Fitch graph. As we shall see below, the placement of HGT
edges in $T$ and the orientation of the edges of the Fitch graph are very
closely related. After introducing the necessary notation in
Section~\ref{sect:prelim}, we give an overview of the concepts and results
in Section~\ref{sec:overview}.

\section{Preliminaries}
\label{sect:prelim}

%% Basics
We denote the \emph{power set} of a set $L$ by $2^L$. A set system
$\mathcal{P}\subseteq 2^L$ is a \emph{partition} of $L$ if \AX{(P0)}
$\emptyset\notin\mathcal{P}$, \AX{(P1)} $\bigcup_{A\in\mathcal{P}} A = L$,
and \AX{(P2)} if $A,B\in\mathcal{P}$ and $A\cap B\ne\emptyset$ then $A=B$.
We write $(x,y)$ for ordered pairs and $\{x,y\}$ for unordered pairs
of elements $x,y\in L$. For a graph $G=(V,E)$, we write $V(G)=V$ and
$E(G)=E$ for its vertex and edge set, respectively. Edges are denoted by
$(x,y)$ in directed graphs and $\{x,y\}$ in undirected graphs.

%% \paragraph{Trees, Hierarchies, and Partitions.}
We consider rooted phylogenetic trees $T$ with root $\rho_T$, vertex set
$V(T)$, edge set $E(T)$, and leaf set $L=L(T)\subseteq V(T)$. The set of
inner vertices is $V^0(T)\coloneqq V(T)\setminus L$ the set of vertices
that are distinct from the leaves.  For any two vertices $x,y\in V(T)$, the
path $P_{x,y}$ connecting $x$ and $y$ in $T$ is a uniquely defined subgraph
of $T$.  The ancestor relation $\preceq_T$ on $V(T)$ is given by
$v\preceq_T u$ if and only if $u$ lies on the unique path from $v$ to the
root.  We write $v\prec_T u$ for $v\preceq_T u$ and $v\ne u$. To
distinguish tree edges from edges of graphs defined on $L$, we write $uv$
instead of $\{u,v\}$ or $(u,v)$ and use the convention that the notation
$uv$ for edges in $T$ implies $v\prec_T u$, i.e., $v$ is a \emph{child} of
$u$ and $u$ is the unique \emph{parent} of $v$, in symbols
$u=\parent_T(v)$. The set of children of a vertex $u$ is denoted by
$\child_T(u)$.  Two vertices $u,v\in V(T)$ are \emph{comparable} if
$v\preceq_T u$ or $u\preceq_T v$.  The \emph{last common ancestor} of a set
$A\subseteq V(T)$, denoted by $\lca_T(A)$, is the unique
$\preceq_T$-minimal element in $V(T)$ that satisfies $v\preceq_T \lca_T(A)$
for all $v\in A$.  For simplicity, we often write $\lca_T(x,y)$ instead of
$\lca_T(\{x,y\})$.  We will often make use of the observation that
$\lca_{T}(A\cup B) = \lca_{T}(\lca_{T}(A),\lca_{T}(B))$ for two subsets
$A,B\subseteq V(T)$.  For $u\in V(T)$, we denote by $T(u)$ the subtree of
$T$ induced by $\{v \in V(T) \mid v\preceq_T u\}$ with root $u$.  If there
is no risk of confusion and the context is clear, we will omit the explicit
reference to $T$, that is, we omit the subscript ``$_T$''.

We note that the trees considered here are slightly less general than the
so-called $X$-trees commonly used in mathematical phylogenetics
\cite{Semple:03}. In an $X$-tree, a set of ``taxa'' $X$ is mapped (not
necessarily injectively) to the vertex set $V(T)$ of a tree $T$. Here
we consider phylogenetic trees in which distinct taxa are represented by
distinct leaves of $T$. That is, our phylogenetic trees are equivalent to
$X$-trees in which the taxa set $X$ is mapped bijectively to $L(T)$.

A hierarchy on $L$ is a system $\mathcal{H}\subset 2^L$ of non-empty sets
such that (i) $A\cap B\in \{\emptyset,A,B\}$ for all $A,B\in \mathcal{H}$,
(ii) $L\in\mathcal{H}$, and (iii) $\{x\}\in \mathcal{H}$ for all $x\in L$.
\begin{proposition}\thmref{\cite{Semple:03}}
  \label{prop:1-1}
  Let $\mathcal H$ be a collection of non-empty subsets of $X$.
  Then, $\mathcal{H}$ is a hierarchy on $X$ if and only if
  there is a rooted phylogenetic tree $T$ on $X$ with
  $\mathcal{H} = \{L(T(v)) \mid c \in V (T)\}$.
\end{proposition}
As a consequence, there is a 1-to-1 correspondence between phylogenetic
trees with leaf set $L$ and hierarchies on $L$ by virtue of
$A\in\mathcal{H}$ if and only if $A=L(T(v))$ for $v=\lca_T(A)$.  We often
write $\mathcal{H}(T) \coloneqq \{L(T(v)) \mid v\in V(T)\}$ for the
hierarchy that is associated by $T$ and refer to the sets $L(T(v))$ as
\emph{clusters} in $T$.  In particular, it holds that
$L(T(v))\subseteq L(T(u))$ if and only if $v\preceq_T u$ for all
$u,v\in V(T)$.  The map $\cl_{T}\colon 2^L\to 2^L$ defined by
$\cl_{T}(A)\coloneqq L(T(\lca_T(A)))$, which assigns to $A$ the unique
inclusion-minimal superset in $\mathcal{H}$, is the canonical closure on
$\mathcal{H}$, as it coincides for hierarchies with the intersection of all
$B\in\mathcal{H}$ with $A\subseteq B$.  A tree $T'$ is a \emph{refinement}
of a tree $T$, if $\mathcal{H}(T) \subseteq \mathcal{H}(T')$.
Equivalently, $T'$ is a refinement of $T$, if $T$ can be obtained from $T'$
by a series of edge-contractions.

\emph{Throughout, we will assume that we are given a pair of a tree $T$
  with leaf set $L$ (or its associated hierarchy $\mathcal{H}$) and a
  partition $\mathcal{P}$ of $L$.}

\section{Main Ideas and Results}
\label{sec:overview}

\begin{figure}
  \begin{center}
    \includegraphics[width=0.75\textwidth]{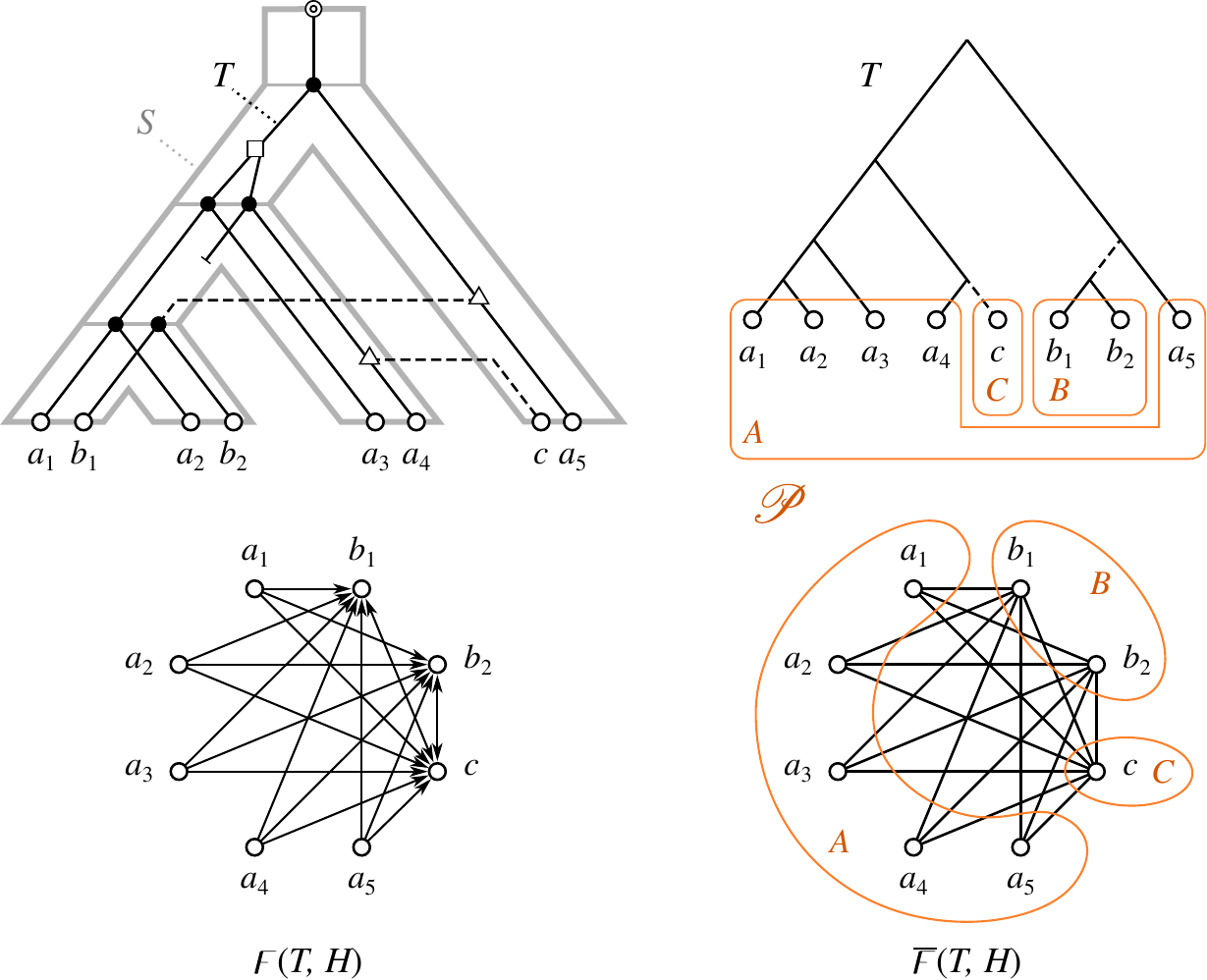}
  \end{center}
  \caption{A gene family history (GFH) (upper left) consisting of a gene
    tree $T$ embedded into a species tree $S$. Apart from speciations
    ($\SPEC$), events such as gene duplications ($\DUPL$), losses
    ($\LOSS$), and horizontal gene transfers ($\HGT$ and dashed lines)
    shape the history of a gene family.  The gene tree $T$ is shown again
    in the upper right panel, without the event labels at the inner
    vertices; the set $H\subseteq E(T)$ of HGT edges is again displayed by
    dashed lines. Removal of $H$ induces the partition
    $\mathcal{P}\coloneqq \partTH(T,H)=\{A,B,C\}$ of $L$.  The directed and
    symmetrized Fitch graphs, $\fitch(T,H)$ and $\symfitch(T,H)$, resp.,
    are shown in the two lower panels. The set of inclusion-maximal
    independent sets of $\symfitch(T,H)$ coincides with $\mathcal{P}$.}
  \label{fig:scenario}
\end{figure}

In the formal representations of gene family histories, an HGT event is a
property of an edge $uv$ in the gene tree $T$. More precisely, $uv$ is an
HGT edge if $u$ and $v$ are mapped into the species tree $S$ in such a
way that that their images are not in an ancestor-descendant relationship,
see Fig.~\ref{fig:scenario} for examples.  Here, we will not
consider GFHs explicitly. Instead, we focus only on gene trees and HGT
events. In \cite{Geiss:18a,Hellmuth:18a} the latter are modeled as an
edge labeling $\lambda: E(T)\to\{0,1\}$. Here, it will be more
convenient to think of HGTs as the subset $H\subseteq E(T)$ of edges in
the gene tree $T$ with $e\in H$ if and only if $\lambda(e)=1$.

Given $T$ and $H\subseteq E(T)$, we write
$T-H \coloneqq (V(T), E(T)\setminus H)$ for the forest obtained by removing
the edges $H$ in $T$.  Moreover, $\partTH(T,H)$ is the partition of $L$
induced by the connected components of $T-H$.  Two genes $x,y\in L$
are called \emph{xenologs} if their history since their last common
ancestor involves a horizontal transfer \cite{Fitch:00}, i.e., if the
path connecting $x$ and $y$ in $T$ contains a transfer edge $e\in
H$. Hence, $x$ and $y$ are xenologs if and only if they are located in
distinct sets of $\partTH(T,H)$.

Xenology has been modeled both with the help of directed and
undirected graphs.  The (directed) \emph{Fitch graph}
$\fitch\coloneqq \fitch(T,H)$ has vertex set $V(\fitch)=L(T)$ and a
directed arc $(x,y)\in E(\fitch)$ if and only if the path $P_{\lca(x,y),y}$
from $\lca_T(x,y)$ to $y$ contains an edge $e\in H$ \cite{Geiss:18a}.  A
digraph $G=(V,E)$ is a Fitch graph if there is a tree $T$ and a subset
$H\subseteq E(T)$ such that $G=\fitch(T,H)$. The \emph{symmetrized Fitch
  graph} $\symfitch\coloneqq\symfitch(T,H)$ has an undirected edge
$\{x,y\}\in E(\symfitch)$ whenever $(x,y)\in E(\fitch)$ or $(y,x)\in E(\fitch)$
\cite{Hellmuth:18a}.  Symmetrized Fitch graphs are complete multipartite
graphs. More precisely, the inclusion-maximal independent sets of
$\symfitch$ form a unique partition $\mathcal{P}$ of $V(G)$ such that
$\{x,y\}\in E(\symfitch)$ if and only if $x$ and $y$ are in distinct sets
of $\mathcal{P}$.  The partition $\partTH(T,H)$ therefore coincides
with the set $\mathcal{P}$ of inclusion-maximal independent sets of
$\symfitch(T,H)$, see Fig.~\ref{fig:scenario}.

Due to the one-to-one correspondence between symmetrized Fitch graphs and 
partitions, it suffices to consider compatibility of trees and partitions.
\begin{definition}
  Let $\mathcal{P}$ be a partition of $L$ and let $T$ be a tree with leaf
  set $L$.  Then $\mathcal{P}$ and $T$ (or equivalently its associated
  hierarchy $\mathcal{H}$) are \emph{compatible} if there is a set
  $H\subseteq E(T)$ such that $\mathcal{P}=\partTH(T,H)$.
  \newline
  In this case, we call $H$ a \emph{separating set} (for $(T,\mathcal{P})$)
  and say that $T$ (and, equivalently, $\mathcal{H}$) and $\mathcal{P}$ are
  \emph{$H$-compatible}.
  \label{def:compatible-rooted}
\end{definition}
Equivalently, one may define $T$ and $\symfitch$ as compatible if
$\symfitch=\symfitch(T,H)$ for some subset $H\subseteq E(T)$.
\citet[Thm.~4.5]{Hellmuth:21q} showed that $T$ and $\mathcal{P}$ are
compatible if and only if, for all $A,B\in\mathcal{P}$, it holds (i)
$\cl_{T}(A)$ is a union of sets of $\mathcal{P}$ and (ii)
$\cl_{T}(A)=\cl_{T}(B)$ implies $A=B$.

Since $\symfitch(T,H)$ is a complete multipartite graph whose independent
sets are formed precisely by the vertices of the parts in $\mathcal{P}$,
the quotient graph $\symfitch(T,H)/\mathcal{P}\simeq K_{|\mathcal{P}|}$ is a 
complete graph on $|\mathcal{P}|$ vertices. 

\begin{figure}[t]
  \begin{center}
\includegraphics[width=0.7\textwidth]{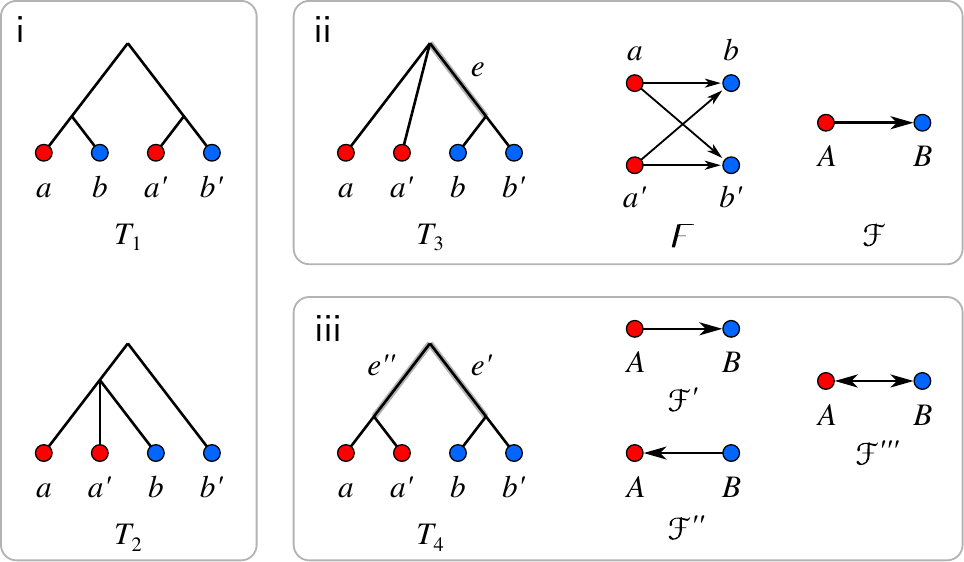}
  \end{center}
  \caption{Examples of trees that are (in)compatible with the partition
    $\mathcal{P}=\{A\coloneqq\{a,a'\},B\coloneqq\{b,b'\}\}$. (i) Both $T_1$
    and $T_2$ are incompatible with $\mathcal{P}$ since every edge in $T$
    lies on a path that connects elements from $A$ or $B$.  (ii) The tree
    $T_3$ is compatible with $\mathcal{P}$ with a unique choice for the
    separating set $H=\{e\}$. Hence, the only possible Fitch graph is
    $\fitch=\fitch(T,H)$ with corresponding Fitch quotient graph
    $\qfitch = \fitch/\mathcal{P}$.  (iii) The tree $T_4$ is compatible with
    $\mathcal{P}$ and admits three valid choices for the separating set
    $H'=\{e'\}$, $H''=\{e''\}$, and $H'''=\{e',e''\}$ yielding Fitch
    quotient graphs $\qfitch'$, $\qfitch''$, and $\qfitch'''$,
    respectively.}
  \label{fig:qfitch-example}
\end{figure}

We will show that the directed quotient digraph
$\qfitch(T,H)\coloneqq \fitch(T,H)/\mathcal{P}$ completely determines the
edges of $\fitch(T,H)$ such that $(x,y)\in E(\fitch(T,H))$ if and only if
$x\in A$, $y\in B$ and $(A,B)\in E(\qfitch(T,H))$, see
Lemma~\ref{lem:quotient-graph} and Cor.~\ref{cor:edges} below.  Therefore,
it suffices to consider $\qfitch(T,H)$, which we will occasionally denote
by $\qfitch$ for brevity. Since its symmetrized version is a complete
graph, we have $(A,B)\in E(\qfitch(T,H))$ or $(B,A)\in E(\qfitch(T,H))$ for
any two distinct $A,B\in\mathcal{P}$. An example for a Fitch graph $\fitch$
and its corresponding quotient graph $\qfitch$ is shown in
Fig.~\ref{fig:qfitch-example}(ii).

\begin{definition}\label{def:Fquotient}
  A graph $G$ is a \emph{Fitch quotient graph} for $(T,\mathcal{P})$ if
  there is $H\subseteq E(T)$ such that $G=\qfitch(T,H)$.
\end{definition}

Below, we compile the notation for the different variants of Fitch graphs
and associated constructs appearing throughout this contribution:
\begin{center}\renewcommand{\arraystretch}{1.2} %Zeilenabstand anpassen
  \begin{tabular}{lll}
%    \hline
    $\fitch\coloneqq \fitch(T,H)$ & \dots & (directed) Fitch graph \\
    $\qfitch\coloneqq \qfitch(T,H)\coloneqq \fitch(T,H)/\mathcal{P}$ & \dots &
    (directed) Fitch quotient graph\\
    $\symfitch\coloneqq\symfitch(T,H)$ & \dots & symmetrized Fitch graph \\
    $\symfitch(T,H)/\mathcal{P}\simeq K_{|\mathcal{P}|}$ & \dots & quotient of
    symmetrized Fitch graph\\
    $\partTH(T,H)$ & \dots & partition of $L$ based on  $T-H$ \\
    $\mathcal{P}$ & \dots & arbitrary partition of $L$\\
    $\mathcal{H}$ & \dots & arbitrary hierarchy on $L$\\
%    \hline
  \end{tabular}
\end{center}

\smallskip In Section~\ref{sect:H}, we will be concerned with the question
to what extent the separating set $H$ is already determined by $T$ and
$\mathcal{P}$.
\begin{definition}
  \label{def:tree-edges}
  An edge $h\in E(T)$ is an \emph{essential separating edge} if $h\in H$
  for all $H$ satisfying $\partTH(T,H)=\mathcal{P}$, a \emph{forbidden
    separating edge} if $h\in H$ for all $H$ satisfying
  $\partTH(T,H)=\mathcal{P}$, and an \emph{ambiguous separating edge}
  otherwise.
\end{definition}
Consider the examples in Fig.~\ref{fig:qfitch-example}. The edge $e$ in the
tree $T_3$ is clearly an essential separating edge since it is the only
edge that separates the set $A,B\in\mathcal{P}$ without at the same time
separating elements from $A$ or $B$. In contrast, the edges on the path
connecting $a$ and $a'$ (or $b$ and $b'$) are clearly forbidden separating
edges.  Finally, the edges $e'$ and $e''$ in $T_4$ are ambiguous separating
edges since $H'=\{e'\}$ and $H''=\{e''\}$ are both valid choices for the
separating set. In order to classify tree edges as (un)ambiguously present
or absent in the separating set, we will use the following vertex coloring,
which is well-defined for a compatible pair $(T,\mathcal{P})$ as a
consequence of Lemma~\ref{lem:path} below.
\begin{definition}
  \label{def:vertex-coloring}
  Let $T$ and $\mathcal{P}$ be compatible. Then,
  $\varpi\colon V(T)\to \mathcal{P} \cup \{\emptyset\}$ is defined by
  putting, for all $v\in V(T)$, $\varpi(v)\coloneqq A$ if there are
  $x,y\in A$ such that $v$ lies along the path connecting $x$ and $y$ in
  $T$. If no such path exists, then $\varpi(v)\coloneqq \emptyset$.
\end{definition}
We will refer to $u\in V(T)$ as a \emph{colored vertex} if
$\varpi(u)\ne\emptyset$. For all $x,y\in A$ and all $A\in\mathcal{P}$ we
have, by construction, $\varpi(x)=\varpi(y)=A$, $\varpi(\lca(x,y))=A$, and
$\varpi(\lca(A))=A$.
The main result of Section~\ref{sect:H} characterizes the classification 
of tree edges in Def.~\ref{def:tree-edges} in terms of the vertex coloring 
$\varpi$:
\smallskip\par\noindent\textbf{Main Result A (Thm.~\ref{thm:tree-edges}).}
An edge $vw\in E(T)$ is an essential separating edge if and only if
$\emptyset\ne\varpi(v)\ne\varpi(w)\ne\emptyset$;
a forbidden separating edge if and only if $\emptyset\ne\varpi(v)=\varpi(w)$;
and an ambiguous separating edge otherwise.
\smallskip \par\noindent
A particular separating set $H$ for a compatible pair $(T,\mathcal{P})$
also determines whether or not a pair
$(A,B)\in \mathcal{P}\times\mathcal{P}$ is an edge of
$\qfitch(T,H)$. Naturally, we ask to what extent the presence or absence of
an edge $(A,B)$ is already determined by $T$ and $\mathcal{P}$.  The
examples in Fig.~\ref{fig:qfitch-example}(ii) and~(iii), respectively, show
that, for a given compatible pair of a partition $\mathcal{P}$ and a tree
$T$, edges in the Fitch quotient graph may be unambiguously present (or
absent) or only present for specific choices of the separating set.

\begin{definition}
  Let $T$ and $\mathcal{P}$ be compatible and let $A,B\in\mathcal{P}$ be
  distinct. Then $(A,B)$ is \emph{essential} if $(A,B)\in E(\qfitch(T,H))$
  for all separating sets $H$ of $(T,\mathcal{P})$, and $(A,B)$ is
  \emph{forbidden} if $(A,B)\notin E(\qfitch(T,H))$ for every separating
  set $H$ of $(T,\mathcal{P})$.
  Otherwise, we say that $(A,B)$ is \emph{ambiguous}.
\end{definition}
In particular, if $(A,B)$ is ambiguous, then there are choices of
separating sets $H_1$ and $H_2$ for $(T,\mathcal{P})$ such that
$(A,B)\in E(\qfitch(T,H_1))$ and $(A,B)\notin E(\qfitch(T,H_2))$.

In Section~\ref{sect:FitchQ}, we turn to characterizing essential and
forbidden edges in the Fitch quotient graphs for compatible $T$ and
$\mathcal{P}$. The main result of this section provides a complete
characterization of $(A,B)$ as essential, forbidden, or ambiguous in terms
of the vertex coloring $\varpi$ of $T$ and the relative positions of the
last common ancestors of $A,B\in\mathcal{P}$, and $A\cup B$ in $T$. More
precisely, we will show:
\smallskip\par\noindent\textbf{Main Result B (Thm.~\ref{thm:direction}).}
$(A,B)$ is essential if and only if there is a colored vertex $v\in V^0(T)$
such that $\lca(B)\prec v\preceq \lca(A\cup B)$; $(A,B)$ is forbidden if
and only if $\lca(A)\prec\lca(B)$; and $(A,B)$ is ambiguous otherwise.
\smallskip \par\noindent
Furthermore, we describe an $O(|L|+|\mathcal{P}|^2)$ algorithm that
computes this classification explicitly for all pairs $A,B\in\mathcal{P}$,
see Cor.~\ref{cor:algo}.

So far, we have assumed that $T$ comprises all (not necessarily binary)
branching events that occurred in the evolution of the gene
family. However, poorly supported edges are often contracted in
phylogenetic reconstructions.  Combinatorial approaches for gene tree
reconstruction also typically only infer a minor of the gene tree $T$.  It
is therefore of practical interest to consider also possible refinements
$T'$ of $T$. More precisely, we ask in Section~\ref{sect:refinements}
whether there are arcs $(x,y)$ that are present (or absent) in the Fitch
graph $\fitch(T',H)$ for every refinement $T'$ of $T$ that is compatible
with $\mathcal{P}$ and every separating set $H$ for $(T',\mathcal{P})$.  It
again suffices to consider the Fitch quotient graphs (cf.\
Lemma~\ref{lem:quotient-graph} and Cor.~\ref{cor:edges}).

\begin{figure}[t]
  \begin{center}
    \includegraphics[width=0.7\textwidth]{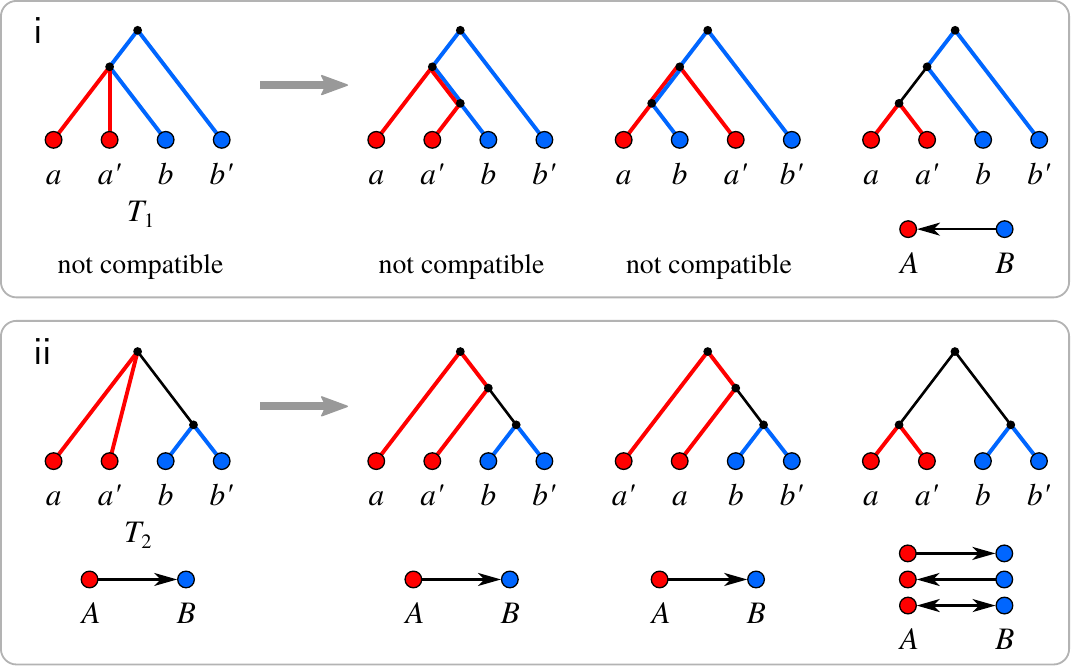}
  \end{center}
  \caption{Two trees $T_1$ and $T_2$ that are r-compatible with the
    partition
    $\mathcal{P}\coloneqq\{A\coloneqq\{a,a'\}, B\coloneqq\{b,b'\}\}$.  Edge
    colors represent $\gamma$ (black edges have color $\emptyset$), except
    in the first two refinements of $T_1$ since these are not r-compatible
    and thus $\gamma$ is not defined.  (i) The tree $T_1$ is not compatible
    with the partition $\mathcal{P}$.  It admits three refinements. Since
    only the third refinement is compatible with $(B,A)$ being essential
    and $(A,B)$ being forbidden, we have that $(B,A)$ is r-essential and
    $(A,B)$ is r-forbidden.  (ii) The tree $T_2$ is compatible with
    $\mathcal{P}$ with $(A,B)$ being essential and $(B,A)$ being
    forbidden. Its refinement on the r.h.s.\ is the example from
    Fig.~\ref{fig:qfitch-example}(iii) admitting multiple choices of the
    separating set. As a consequence, both $(A,B)$ and $(B,A)$ are
    r-ambiguous.}
  \label{fig:refinement-examples}
\end{figure}

\begin{definition}\label{def:r-comp}
  A tree $T$ with leaf set $L$ is \emph{refinement-compatible}
  (\emph{r-compatible} for short)  with a
  partition $\mathcal{P}$ of $L$ if there is a refinement $T'$ of $T$ that
  is compatible with $\mathcal{P}$.
\end{definition}
Note that a tree $T$ can be r-compatible with $\mathcal{P}$ although $T$
and $\mathcal{P}$ are not compatible, see
Fig.~\ref{fig:refinement-examples}(i) for an example.  Answering the
question whether $T$ is r-compatible with $\mathcal{P}$ thus provides
additional information about the topology of phylogenetic trees that is
implicitly contained in $\mathcal{P}$ or $\symfitch$.  Refinement
compatibility was characterized by \citet{Hellmuth:21q} in the following
manner:
\begin{proposition}\thmref{\cite[Prop.~7.3 and Thm.~7.5]{Hellmuth:21q}}
  \label{prop:refiment-existence}
  Let $T$ be a tree on $L$ and $\mathcal{P}$ be a partition of $L$. Then
  $T$ and $\mathcal{P}$ are r-compatible if and only if there is no edge
  $e\in E(T)$ lying on the path from $x$ to $x'$ and on the path from $y$
  to $y'$ such that $x,x'\in A$ and $y,y'\in B$ for distinct
  $A,B\in \mathcal{P}$.  In the positive case, a compatible refinement $T'$
  of $T$ can be constructed in $O(|L|)$ operations.
\end{proposition}
Moreover, \citet[Prop.~5.1]{Hellmuth:21q} showed that compatibility of $T$
and $\mathcal{P}$ implies that all refinements of $T$ are again compatible
with $\mathcal{P}$. The vertex coloring $\varpi$ is in general not
well-defined for an r-compatible pair $(T,\mathcal{P})$ because $T$ does
not need to be compatible with $\mathcal{P}$.  To see this, consider again
Fig.~\ref{fig:refinement-examples}(i) and the common parent of vertices
$a$, $a'$, and $b$ in $T_1$. This vertex lies on the path connecting
$a,a'\in A$, as well as on the path connecting $b,b'\in B$.  Instead of
$\varpi$, we therefore consider an edge coloring similar to the one used in
\cite{Hellmuth:21q}:
\begin{definition}
  Let $T$ and $\mathcal{P}$ be r-compatible. Then,
  $\gamma\colon E(T)\to \mathcal{P} \cup \{\emptyset\}$ is defined by
  putting, for all $e\in E(T)$, $\gamma(e)\coloneqq A$ if there are
  $x,y\in A$ such that $e$ lies along the path connecting $x$ and $y$ in
  $T$, in which case we say that $e$ is \emph{colored}.
  If no such path exists, then $\gamma(e)\coloneqq \emptyset$.
\end{definition}
The only difference to the edge coloring in \cite{Hellmuth:21q} is that, as
a consequence of Prop.~\ref{prop:refiment-existence}, we can directly use
the sets $A\in \mathcal{P}$ as ``colors'' rather than subsets of
$2^{\mathcal{P}}$. Generalizing the notion of essential edges in Fitch
quotient graphs, we consider pairs $(A,B)$:
\begin{definition}
  Let $T$ and $\mathcal{P}$ be r-compatible and let $A,B\in\mathcal{P}$ be
  distinct. We say that $(A,B)$ is \emph{r-essential} if
  $(A,B)\in \qfitch(T',H)$ for all separating sets $H\subseteq E(T')$ of
  every refinement $T'$ of $T$ that is compatible with $\mathcal{P}$. We
  say that $(A,B)$ is \emph{r-forbidden} if $(A,B)\notin \qfitch(T',H)$ for
  any separating set $H\subseteq E(T')$ of any refinement $T'$ of $T$ that
  is compatible with $\mathcal{P}$.  In all other cases, $(A,B)$ is
  \emph{r-ambiguous}.
\end{definition}

If $T$ and $\mathcal{P}$ are already compatible and $(A,B)$ is ambiguous
for two sets $A,B\in\mathcal{P}$, then clearly $(A,B)$ is also r-ambiguous
since $T$ is a compatible refinement of itself. Analogous statements,
however, do not hold for essential and forbidden edges as the example in
Fig.~\ref{fig:refinement-examples}(ii) shows. By definition, an essential
edge $(A,B)$ cannot be r-forbidden, and a forbidden edge $(A,B)$ cannot be
r-essential.  Therefore, considering refinements cannot decrease the level of 
ambiguity.

The main result of Section~\ref{sect:refinements} is a characterization of
r-essential and r-forbidden pairs $(A,B)$ for a given a tree $T$ and an
r-compatible partition $\mathcal{P}$. Similar to our first two main
results, the characterization can be expressed in terms of the edge
coloring $\gamma$ and the last common ancestors of $A$, $B$, and
$u\coloneqq\lca_T(A\cup B)$ in $T$:
\smallskip\par\noindent\textbf{Main Result C 
(Thm.~\ref{thm:direction-refinement}).}
$(A,B)$ is r-essential if and only if (a) the path $P_{u, \lca_{T}(B)}$
contains a colored edge, or (b) $u\ne\rho_T$ and the path $P_{u,\lca_{T}(A)}$
contains a colored edge $e$ with
$\gamma(e)=\gamma(\parent_T(u) u)=C\in\mathcal{P}$.
$(A,B)$ is r-forbidden if and only if $\lca_{T}(A) \prec_T \lca_{T}(B)=u$
and $\gamma(uv)=B$ for the vertex $v\in \child_{T}(u)$ satisfying
$\lca_{T}(A)\preceq_T v$.
\smallskip \par
\begin{figure}[t]
  \begin{center}
    \includegraphics[width=0.6\textwidth]{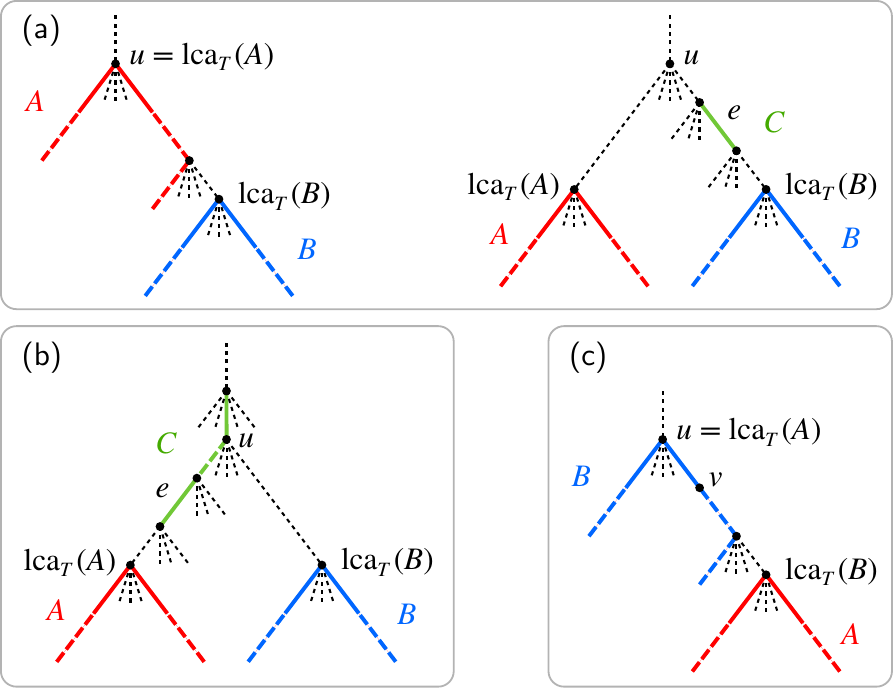}
  \end{center}
  \caption{The situations leading to r-essential or r-forbidden pairs
    $(A,B)$ with $A,B\in\mathcal{P}$ and $u\coloneqq \lca_{T}(A\cup
    B)$. Dashed lines indicate paths that may or may not exist. (a) $(A,B)$
    is r-essential if the path $P_{u, \lca_{T}(B)}$ contains a colored
    edge. This edge may have color $A$ (left) in which case $u=\lca_{T}(A)$
    or a color $C\in\mathcal{P}\setminus \{A,B\}$ (right).  (b) The second
    possibility leading to $(A,B)$ being r-essential.  (c) The situation
    when $(A,B)$ is r-forbidden.}
  \label{fig:refinement-thm}
\end{figure}

\noindent
The different situations that make $(A,B)$ r-essential or r-forbidden are
illustrated in Fig.~\ref{fig:refinement-thm}.
The characterization also gives rise to an $O(|L|+|\mathcal{P}|^2)$
algorithm that computes this classification explicitly for all pairs
$A,B\in\mathcal{P}$.
Implementations of the classification algorithms
based on the three main results are applied to simulated GFHs in
Section~\ref{sect:comp} to determine the relative abundances of essential,
forbidden, and ambiguous edges in the gene trees and the corresponding
Fitch quotient graphs. We will see that ambiguities are surprisingly rare
in these evolutionary scenarios.

\section{Possible Choices of the Separating Set $H$}

\label{sect:H}

In order to understand the ambiguity of edges in the Fitch quotient graph
$\qfitch(T,H)$, we first need to understand to what extent the choice of the
separating set $H$ is constrained by a given tree $T$ and a partition
$\mathcal{P}$.  Our starting point is
\begin{proposition}\thmref{\cite[Cor.~7.7]{Hellmuth:21q}}
  If $T$ and $\mathcal{P}$ are compatible, then there is a unique
  inclusion-maximal separating set $H^*$ such that
  $\mathcal{P}=\partTH(T,H)$.
  \label{prop:uniqH*}
\end{proposition}
By \cite[Thm.~7.6]{Hellmuth:21q}, it can be constructed explicitly as the
set of all edges $e\in E(T)$ that do not lie along the path connecting a
pair of points $x,y\in A$ for any $A\in\mathcal{P}$. Here, we use an even
simpler, vertex-centered construction based on the following simple
observation:
\begin{lemma}
  Let $T$ and $\mathcal{P}$ be compatible, $A,A'\in\mathcal{P}$ and
  $v\in V(T)$. If $v$ lies on both the path $P_{x,y}$ connecting $x,y\in A$
  and the $P_{x',y'}$ connecting $x',y'\in A'$, then $A=A'$.
  \label{lem:path}
\end{lemma}
\begin{proof}
  First we note that $E(P_{x,y})\cap H^*=\emptyset$ and
  $E(P_{x',y'})\cap H^*=\emptyset$.  The union of the path $P_{x,y}$ and
  $P_{x',y'}$ together form a (non-phylogenetic) sub-tree $T_4$ of $T$. We
  therefore have
  $v\in V(P_{x,y})\cap V(P_{x',y'})= V(P_{x,x'})\cap V(P_{y,y'})$ or
  $v\in V(P_{x,y})\cap V(P_{x',y'})= V(P_{x,y'})\cap V(P_{x',y})$ and thus
  $v$ also lies along one of the paths $P_{x,y'}$, $P_{x,x'}$, $P_{x',y}$,
  or $P_{y,y'}$. Together with $E(T_4)\cap H^*=\emptyset$, this implies
  $A\cap A'\ne\emptyset$ and thus $A=A'$ because $\mathcal{P}$ is a
  partition.
\end{proof}

We can therefore uniquely ``color'' each vertex $v\in V(T)$ by a set
$\varpi(v)=A\in\mathcal{P}$ whenever there are $x,y\in A$ such that $v$
lies along the path connecting $x$ and $y$ in $T$; otherwise we set 
$\varpi(v)=\emptyset$. Hence, the vertex coloring $\varpi\colon V(T)\to
\mathcal{P} \cup \{\emptyset\}$ in Def.~\ref{def:vertex-coloring} is
well-defined.
\begin{lemma}
  \label{lem:varpi-compute}
  Let $T$ and $\mathcal{P}$ be $H$-compatible. Then, the map $\varpi$ as
  well as $\lca(A)$ for all $A\in\mathcal{P}$ can be computed in $O(|L|)$
  total time.
\end{lemma}
\begin{proof}
  This proof parallels the construction of an associated edge coloring in
  the proof of Thm.~7.5 in \cite{Hellmuth:21q}.  The LCA data structure
  described by \citet{Bender:05} enables constant time look-up of
  $\lca_T(u,v)$ for any $u,v\in V(T)$ after an $O(|L|)$ preprocessing step.
  We now show how to compute the map $\varpi$ as well as $\lca(A)$ for all
  $A\in\mathcal{P}$ in linear total time.

  We start by initializing $\varpi(v)=\emptyset$ for all $v\in V(T)$ in
  $O(|L|)$.  We then process every $A=\{x_1,\dots,x_k\}\in\mathcal{P}$ as
  follows.  First, we initialize the set of previously visited vertices of
  $V(T)$ as $\texttt{visited}\leftarrow\emptyset$ and the current last
  common ancestor as $\texttt{curLCA}\leftarrow x_1$.  The latter will be
  updated stepwisely such that it equals $\lca_T(A)$ in the end.  To this
  end, for each leaf $x\in\{x_2,\dots,x_k\}$ (if any), we query
  $\texttt{newLCA}=\lca_T(x,\texttt{curLCA})$ and move from $x$ upwards
  along the tree. We set $\varpi(v)=A$ for each vertex $v$ encountered
  during the traversal, and add $v$ to \texttt{visited}.  The traversal
  stops as soon as $v$ is in \texttt{visited} or equals $\texttt{newLCA}$.
  In case we have $\texttt{curLCA}\prec_T\texttt{newLCA}$, which by
  definition of \texttt{newLCA} holds if
  $\texttt{curLCA}\ne\texttt{newLCA}$, we perform the same bottom-up
  traversal starting from $\texttt{curLCA}$.  As a final step in the
  processing of $x$, we set $\texttt{curLCA}\leftarrow\texttt{newLCA}$.
  One easily verifies that, after processing all vertices in $A$, it holds
  $\lca(A)=\texttt{curLCA}$ and we have exactly colored the vertices in the
  minimal subtree of $T$ that connects all leaves in $A$, i.e., the
  vertices that lie on a path connecting two $x,x'\in A$.  Moreover, each
  vertex considered in the bottom-up traversals is colored with $A$ and
  required only a constant number of constant-time queries and operations.
  As a consequence of Lemma~\ref{lem:path}, no vertex is colored a second
  time in a subsequent bottom-up traversal for another set in
  $\mathcal{P}$.  A vertex is considered twice when processing $A$ only if
  it is already in \texttt{visited}. This occurs at most twice per vertex
  $x\in L$ (the last vertex in the first traversal starting from $x$ and
  the first vertex in the second traversal starting at
  $\texttt{curLCA}$). Thus, the overall effort for vertices encountered
  more than once is bounded by $O(|L|)$.  The additional operations needed
  for each $x\in A$ (i.e., set initialization, query, comparison, and
  update of the last common ancestor) are performed in constant time.
  Since $T$ is a phylogenetic rooted tree, we have $|V(T)|\le 2|L|-1$. In
  total, therefore, the traversals of the tree require $O(|L|)$
  operations. In addition, a constant effort is required for each of the
  $O(|L|)$ vertices in the disjoint sets in $\mathcal{P}$.  In summary, we
  have computed $\varpi$ as well as $\lca(A)$ for all $A\in\mathcal{P}$ in
  $O(|L|)$ total time.
\end{proof}

Lemma~\ref{lem:path} implies that an edge $vw\in E(T)$ either connects two
colored vertices with $\varpi(v)=\varpi(w)$, in which case $vw\notin H^*$,
or it does not lie along any path connecting leaves $x,y\in A$ for some
$A\in\mathcal{P}$ and thus $vw\in H^*$. We therefore can characterize $H^*$
in terms of the vertex coloring $\varpi$ as follows:
\begin{fact}
  \label{obs:H*}
  For all edges $vw\in E(T)$ it holds that $vw\notin H^*$ if and only if
  $\varpi(v)=\varpi(w)\ne\emptyset$. In particular, if
  $\varpi(v)=\emptyset$ then all edges $e$ incident with $v$ satisfy
  $e\in H^*$.
\end{fact}
The following observation plays a key role in our analysis, see also
\cite[Cor.~7.7]{Hellmuth:21q}:
\begin{lemma}
  \label{lem:HsubH*}
  Let $\mathcal{P}$ be a partition of $L$. Then for every
  $H\subseteq E(T)$, $\partTH(T,H)=\mathcal{P}$ implies $H\subseteq H^*$.
\end{lemma}
\begin{proof}
  Suppose, for contradiction, that $\partTH(T,H)=\mathcal{P}$ and that
  there is an edge $h\in H\setminus H^*$, i.e., it holds $h=vw$ with
  $\varpi(v)=\varpi(w)\ne\emptyset$ by Obs.~\ref{obs:H*}.  By definition of
  $\varpi$, thus, $h$ lies along the (unique) path $P_{x,y}$ between two
  leaves $x,y\in L(T)$ that belong to same subset
  $\varpi(x)=\varpi(y)=\varpi(v)=\varpi(w)= A \in\mathcal{P}$. However, $x$
  and $y$ are located in different connected components of $T-H$, and thus
  $A\notin \partTH(T,H)$. Hence, we have $\mathcal{P}\ne \partTH(T,H)$; a
  contradiction.
\end{proof}
By Obs.~\ref{obs:H*}, Lemma~\ref{lem:HsubH*} and since $H^*$ always is a 
separating set for a compatible pair $(T,\mathcal{P})$, we obtain:
\begin{corollary}
  \label{cor:forbidden-HGT}
  An edge $vw\in E(T)$ is a forbidden separating edge if and only if
  $vw\notin H^*$ if and only if $\varpi(v)=\varpi(w)\ne\emptyset$.
\end{corollary}

\begin{lemma}
  \label{lem:H-edge-distinct-colors}
  Let $T$ and $\mathcal{P}$ be compatible and $u,v\in V(T)$ be two colored
  vertices.  Then, for any choice of $H$, $\varpi(u)\ne\varpi(v)$ if and
  only if $E(P_{u,v})\cap H \ne \emptyset$.
\end{lemma}
\begin{proof}
  Let $H$ be an arbitrary separating set such that $T$ and $\mathcal{P}$
  are $H$-compatible, and let $A,B\in \mathcal{P}$ such that $\varpi(u)=A$
  and $\varpi(v)=B$.  By definition, therefore, there are paths $P_{a,u}$
  for some $a\in A$ and $P_{b,v}$ for some $b\in B$. In particular, these
  paths do not contain edges in $H$ and all vertices along these paths are
  colored with $A$ and $B$, respectively.

  Assume first that $\varpi(u)\ne\varpi(v)$, i.e., $A,B\in \mathcal{P}$ are
  distinct.  In this case, $P_{a,u}$ and $P_{b,v}$ must be vertex-disjoint.
  Since $T$ is connected, there must be vertices $u'\in V(P_{a,u})$ and
  $v'\in V(P_{a,v})$ such that
  $P_{a,b}=P_{a,u'}\cup P_{u',v'} \cup P_{v',b}$.  At least one of the
  edges in $P_{u',v'}$ must be contained in $H$, since otherwise the path
  $P_{a,b}$ remains in $T-H$ connecting vertices $a$ and $b$ from two
  distinct sets $A,B\in\mathcal{P}$.  Since $u\in V(P_{a,u})$ and
  $v\in V(P_{a,v})$ and paths in trees are unique, we clearly have
  $P_{u',v'}\subseteq P_{u,v}$ and thus $P_{u,v}$ contains an edge in $H$.

  Now assume $E(P_{u,v})\cap H \ne \emptyset$.  Hence, $u$ and $v$ are in
  distinct connected components of the forest $T-H$.  Since $P_{a,u}$ and
  $P_{b,v}$ do not contain edges in $H$, they are still paths in
  $T-H$. Together, the latter two arguments imply that $a\in A$ and
  $b\in B$ also lie in distinct connected components of $T-H$.  Therefore,
  we obtain $\varpi(u)=A\ne B=\varpi(v)$.
\end{proof}
In particular, if two colored vertices $u$ and $v$ are connected by a
single edge $e$, then $e\in H$ if and only if $\varpi(u)\ne\varpi(v)$.

\begin{figure}[t]
  \begin{center}
    \includegraphics[width=0.8\textwidth]{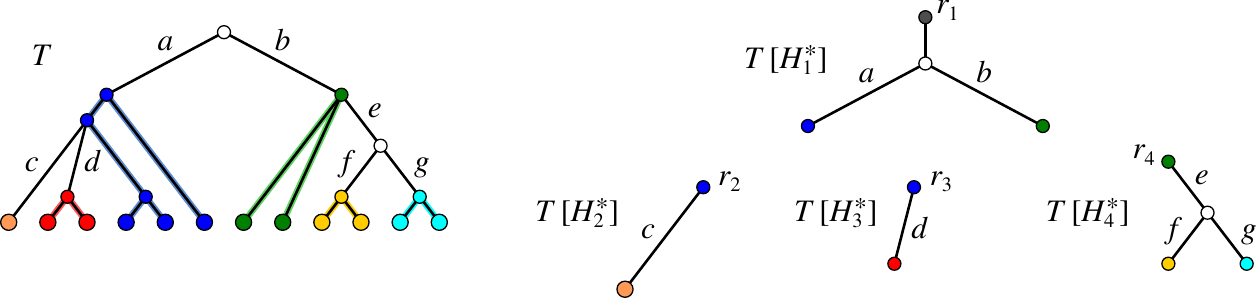}
  \end{center}
  \caption{The tree $T$ is compatible with a partition $\mathcal{P}$ where
    leaves have the same color iff they are in the same set of
    $\mathcal{P}$.  The edges in the unique separating set $H^*$ are
    labeled by lower-case letters and inner vertices $v$ for which
    $\varpi(v)=\emptyset$ are white. The colored vertices partition $H^*$
    into $H_1^*,H_2^*,H_3^*,$ and $H_4^*$. The subtrees $T[H_i^*]$ induced
    by the sets $H_i^*$ are shown on the right side. Note that a ``planted
    rooted'' with a dummy color was added to $T[H_1^*]$.
    }
  \label{fig:H-partition}
\end{figure}

The edge set $H^*$ is naturally partitioned into disjoint components
$H_i^*$ that are separated by colored vertices. More formally, two distinct
edges $uv,u'v'\in H^*$ are in the same set $H_i^*$ if and only if at least
one of the paths $P_{u,u'}$, $P_{u,v'}$, $P_{v,u'}$, and $P_{v,v'}$ does
not contain a colored vertex; see Fig.~\ref{fig:H-partition} for an
example. Note that both $uv$ and $u'v'$ contain at least one uncolored
vertex and no two edges $h,h'\in H_i^*$ are incident with the same colored
vertex $u\in V(T)$.  No uncolored vertex $v\in V(T)$ can be incident to
edges that are contained in distinct sets $H_i^*$ and $H_j^*$. Thus, each
edge set $H_i^*$ therefore defines a subtree
\begin{equation*}
  T[H_i^*] \coloneqq T[\{v\in V(T)\mid v \text{ is incident with some } e\in
  H_i^*\}]
\end{equation*}
All leaves of $T[H_i^*]$ are colored and all inner vertices except possibly the
root, i.e., the $\preceq$-maximal element of $T[H_i^*]$, are uncolored.
The root of $T[H_i^*]$ is either a colored vertex $u$ incident with a single
edge $uw\in H_i^*$, or the uncolored root of $T[H_i^*]$ coincides with the
root $\rho$ of $T$. If the latter special case occurs,
we add a ``planted rooted'' with an additional dummy color
not corresponding to a set of $\mathcal{P}$.
The latter construction allows us to handle the component  $T[H_i^*]$
containing the root of $T$ in the same manner as all other components.
We write $r_i$ for the
root (or the additional planted root in the special case), $L[H_i^*]$ for
the leaf set and $V^0[H_i^*]$ for the set of inner vertices of $T[H_i^*]$
not including $r_i$.  Note that the elements of $L[H_i^*]$ are not
necessarily leaves of $T$.  By construction, $\varpi(u)=\emptyset$ if and
only if $u\in V^0[H_i^*]$. Moreover, we have $\varpi(u)\ne\varpi(v)$ for
any two distinct $u,v\in L[H_i^*]\cup\{r_i\}$. We say that $H_i$ is a
separating set on $T[H_i^*]$ if every path connecting two distinct
leaves or a leaf and the root $r_i$ of $T[H_i^*]$ contains an edge $h\in H_i$.
By way of example, for every choice of $H_4\subseteq \{e,f,g\}$ with
$|H_4|\geq 2$, the set $H_4$ is a separating set on $T[H_4^*]$ in
Fig.~\ref{fig:H-partition}, while $H_4$ cannot be a separating set on
$T[H_4^*]$ as soon as $|H_4|< 2$.
As we shall see in the next result, separating sets $H$ of $T$ can be
characterized in terms of separating sets $H_i$ on the underlying trees
$T[H_i^*]$.

\begin{lemma}
  \label{lem:componentwise}
  Let $T$ and $\mathcal{P}$ be compatible.  $H$ is a separating set for
  $(T,\mathcal{P})$ if and only if $H=\bigcup_i H_i$ such that every
  $H_i\subseteq H_i^*$ is a separating set for $T[H_i^*]$.
\end{lemma}
\begin{proof}
  Let $H$ be a separating set for $(T,\mathcal{P})$ and consider
  $H_i\coloneqq H\cap H_i^*$. If $H_i$ is not a separating set for $T[H_i^*]$
  then there are two colored vertices $a$ and $b$ with
  $\varpi(a)\ne\varpi(b)$ connected by a path without an edge $h\in H$; a
  contradiction to Lemma~\ref{lem:H-edge-distinct-colors}.  Conversely, let
  $H_i$ be a separating set for each component $T[H^*_i]$. Suppose $x\in A$
  and $y\in B$ for distinct $A,B\in \mathcal{P}$. Hence, the path $P_{x,y}$
  contains an edge $h=uv\in H^*$ which is contained in some $H_i^*$. Assume
  w.l.o.g.\ that $u$ is closer to $x$ than $v$. Let $u'$ and $v'$ be the
  last colored vertex on the path from $x$ to $u$ and on the path from $y$
  to $v$, respectively.  By construction, $u'$ and $v'$ are vertices in
  $T[H_i^*]$. In particular, since they are colored, we have
  $u',v'\in L[H_i^*]\cup\{r_i\}$.  Since moreover $H_i$ is a separating set
  for $T[H_i^*]$, the subpath $P_{u',v'}$ of $P_{x,y}$ contains an edge
  $h\in H_i\subseteq H$.  For $x,y\in A$ for some $A\in \mathcal{P}$, the
  path $P_{x,y}$ does not contains an edge $h\in H^*$ and, since
  $H\subseteq H^*$, also not an edge $h\in H$. In summary,
  $H=\bigcup_i H_i$ is separating for $(T,\mathcal{P})$.
\end{proof}

Lemma~\ref{lem:componentwise} suggests to study the separating sets $H_i$
for $T[H^*_i]$ independently of each other. In particular, $H$ is a
separating set for $(T,\mathcal{P})$ if and only if every $H_i=H\cap H_i^*$
is a separating set for $T[H^*_i]$. We immediately observe that $h$ is an
essential separating edge whenever $\{h\}=H^*_i$ for some component of
$H^*$. This, in fact, characterizes the existence of essential separating
edges.

\begin{lemma}
  \label{lem:nonessential}
  An edge $h\in H_i^*$ is an essential separating edge if and only if
  $H^*_i = \{h\}$.
\end{lemma}
\begin{proof}
  If $H^*_i = \{h\}$, then $h$ must be contained in every separating set
  $H_i$ for $T[H^*_i]$ since otherwise there is no edge in $H_i$ between
  the root and the single leaf of $T[H^*_i]$, i.e., the two vertices
  adjacent to $h$. By Lemma~\ref{lem:componentwise}, therefore, $h$ must be
  contained in every $H$ satisfying $\partTH(T,H)=\mathcal{P}$.  Hence, $h$
  is an essential separating edge.

  By contraposition, assume now that $H^*_i \neq \{h\}$ and thus,
  $|H_i^*|>1$.  First we note that in this case $H_i^*$ contains at least
  three edges, two leaves, and one uncolored (interior) vertex. Denote by
  $H_L\subseteq H_i^*$ the set of edges $e_v=\{\parent(v),v\}$ connecting
  the leaves $v\in L[H_i^*]$ of $T[H_i^*]$ to their parents. Clearly $H_L$
  separates all leaves and the root of $T[H_i^*]$ from each other. Thus
  none of the edges that are not incident to a leaf is an essential
  separating edge, and in particular, the single edge $e=r_iu_i$, where
  $u_i$ is the unique child of $r_i$ in $T[H_i]$, is not an essential
  separating edge. On the other hand, for some $v\in L[H_i^*]$ consider the
  set $H_v$ comprising all edges of $H_i^*$ except the edges along the path
  from $v$ to unique child $u_i$ of the root. Clearly, $H_v$ also separates
  all leaves and the root of $T[H_i^*]$ from each other. Therefore $e_v$ is
  not an essential separating edge for all $v\in L[H_i^*]$.  Consequently,
  $H_i^*$ does not contain any essential separating edge.
\end{proof}
We summarize the results of this section in
\begin{theorem}
  \label{thm:tree-edges}
  Let $T$ and $\mathcal{P}$ be compatible. An edge $vw\in E(T)$ is
  \begin{description}
    \item[(1)] an essential separating edge if and only if
    $\emptyset\ne\varpi(v)\ne\varpi(w)\ne\emptyset$,
    \item[(2)] a forbidden separating 
    edge if and only if $\emptyset\ne\varpi(v)=\varpi(w)$.
  \end{description}
   and an ambiguous separating edge otherwise.
   In particular, the edges in $T$ can be classified as essential, forbidden, 
   or ambiguous separating edges in $O(|L|)$.
\end{theorem}
\begin{proof}
  As an immediate consequence of Lemma~\ref{lem:nonessential}, an edge
  $vw\in E(T)$ is an essential separating edge if and only if
  $\emptyset\ne\varpi(v)\ne\varpi(w)\ne\emptyset$.  By
  Cor.~\ref{cor:forbidden-HGT}, $vw\in E(T)$ is a forbidden separating edge
  if and only if $\emptyset\ne\varpi(v)=\varpi(w)$.  By definition,
  $vw\in E(T)$ an ambiguous separating edge otherwise.  By
  Lemma~\ref{lem:varpi-compute}, $\varpi$ can be computed in $O(|L|)$ time.
  Since, for each edge, comparing the colors of its endpoints then takes
  constant time, the total effort to classify all $O(|L|)$ edges is bounded
  by $O(|L|)$.
\end{proof}

\section{Edges of the Fitch Quotient Graph}
\label{sect:FitchQ}

In this section, we answer the question whether or not a pair $(x,y)$ with
$x,y\in L$ is unambiguously present (or absent) in the Fitch graph given
the knowledge of $T$ and $\mathcal{P}$.  We first show that the edge set of
Fitch quotient graph $\qfitch(T,H)=\fitch(T,H)/\mathcal{P}$ completely
determines the edges of the Fitch graph $\fitch(T,H)$.
\begin{lemma}
  \label{lem:quotient-graph}
  Let $T$ and $\mathcal{P}$ be $H$-compatible and $A, B\in \mathcal{P}$.
  Then there are $x\in A, y\in B$ with $(x,y)\in E(\fitch(T,H))$ if and
  only if $(x',y')\in E(\fitch(T,H))$ for all $x'\in A, y'\in B$.
\end{lemma}
\begin{proof}
  Suppose there are $x\in A$ and $y\in B$ such that
  $(x,y)\in\fitch(T,H)$, i.e., the path
  $P\coloneqq P_{\lca_T(x,y)y}\subseteq E(T)$ connecting $\lca_T(x,y)$ and
  $y$ contains an edge $e=uv\in H$.  In particular, $A$ and $B$ must be
  distinct, $y\in L(T(v))$ and $x\notin L(T(v))$.  The latter implies
  $x'\notin L(T(v))$ for any $x'\in A$ since otherwise $P_{x,x'}$ contains
  $e$ contradicting $x,x'\in A$.  By similar arguments, $y'\in L(T(v))$
  holds for any $y'\in B$.  Hence, for any $x'\in A$ and $y'\in B$, we
  obtain $u\preceq_T\lca_T(x', y')\eqqcolon w$.  In particular, since
  $y'\preceq_T v\prec_T u \preceq_T w$, the path $P_{w,y'}$ contains $e$,
  and thus $(x',y')\in E(\fitch(T,H))$.  The converse follows trivially
  from the fact that $A$ and $B$ are sets in the partition $\mathcal{P}$
  and thus non-empty.
\end{proof}
It therefore suffices to consider $\qfitch(T,H)$:
%For later reference we summarize this discussion as
\begin{corollary}\label{cor:edges}
  Let $T$ and $\mathcal{P}$ be $H$-compatible, $A, B\in \mathcal{P}$ and
  $x\in A, y\in B$.  Then, $(x,y)\in E(\fitch(T,H))$ if and only if
  $(A,B)\in E(\qfitch(T,H))$.
\end{corollary}
We first derive sufficient 
conditions for essential pairs $(A,B)\in E(\qfitch(T,H))$ and forbidden pairs 
$(A,B)\notin E(\qfitch(T,H))$.
\begin{proposition}
  \label{prop:unambiguous}
  Let $T$ and $\mathcal{P}$ be compatible. Then the following statements hold
  for all distinct $A,B\in\mathcal{P}$:
  \begin{description}
  \item[(i)] If there is a colored vertex $v\in V^0(T)$ such that
    $\lca(B)\prec v\preceq \lca(A\cup B)$, then $(A,B)$ is essential.
  \item[(ii)] If $\lca(A)\prec \lca(B)$, then $(A,B)$ is forbidden.
  \end{description}
\end{proposition}
\begin{proof}
  (i) Since $v\in V^0(T)$ is colored, it holds that
  $\varpi(v)\ne\emptyset$.  By construction, there is a leaf $x\in A$ and
  $y\in B$ such that
  $y\preceq \lca(B)\prec v\preceq \lca(x,y) \eqqcolon u$.  By assumption,
  we have $\lca(B)\prec v$ and thus $\varpi(v)\ne B$.  By
  Lemma~\ref{lem:H-edge-distinct-colors} and since moreover
  $\varpi(y)=B\ne\varpi(v)$, the path $P_{v,y}$, which is a subpath of
  $P_{u,y}$, contains at least one edge $e\in H$ for any choice of
  $H$. Hence, we have $(x,y)\in E(\fitch(T,H))$ and by Cor.~\ref{cor:edges}
  $(A,B)$ is essential.

  (ii) Since $\lca(A)\prec \lca(B)$, we have
  $\varpi(\lca(A))=A\ne B=\varpi(\lca(B))$.  Moreover, there are leaves
  $x\in A$ and $y\in B$ such that
  $x\preceq \lca(A)\prec \lca(x,y) = \lca(B)$. Cor.\ \ref{cor:edges}
  together with $\varpi(\lca(B))=\varpi(y)=B$ implies that the path
  $P_{\lca(B), y} = P_{\lca(x,y), y}$ contains no edge $w\in H$.  Thus
  $(x,y)\notin E(\fitch(T,H))$ for any choice of $H$, which, by
  Cor.~\ref{cor:edges}, implies that $(A,B)$ is forbidden.
\end{proof}
Note that the vertex $\lca(B)$ in Prop.~\ref{prop:unambiguous} must be
colored.  In particular, if Prop.~\ref{prop:unambiguous}(ii) is satisfied,
then $\lca(A)\prec \lca(B) = \lca(A\cup B)$ and thus,
Prop.~\ref{prop:unambiguous}(i) is satisfied by interchanging the role of
$A$ and $B$. In summary, we obtain
\begin{corollary}
  \label{cor:unambiguous}
  Let $T$ and $\mathcal{P}$ be compatible. If $\lca(A)\prec \lca(B)$ for
  two $A,B\in\mathcal{P}$, then $(B,A)$ is essential and $(A,B)$ is
  forbidden.
\end{corollary}

Recall that $\lca(A)=\lca(B)$ cannot occur for distinct
$A,B\in \mathcal{P}$ \cite[Thm.~4.5]{Hellmuth:21q}.  By
Cor.~\ref{cor:unambiguous}, we can unambiguously infer the edges between
$A$ and $B$ in $\qfitch$ whenever $\lca(A)$ and $\lca(B)$ are
comparable. Whenever $\lca(A)$ and $\lca(B)$ are incomparable and there is
no colored vertex $v\in V^0(T)$ such that
$\lca(B)\prec v\preceq \lca(A\cup B)$, however, we cannot make a statement
about the presence of $(A,B)$ in $E(\qfitch)$.

The construction of the separating sets $H_v$ and $H_L$ in the proof of
Lemma~\ref{lem:nonessential} can be used to infer the following statement
of edges in Fitch graphs.
\begin{lemma}
  \label{lem:ambiguous}
  Let $T$ and $\mathcal{P}$ be compatible and $A,B\in\mathcal{P}$ be
  distinct.  If $\lca(A)$ and $\lca(B)$ are $\preceq$-incomparable and
  there is no colored vertex $v$ with
  $\lca(B)\prec v \preceq \lca(A\cup B)$, then $(A,B)$ is
  ambiguous.
\end{lemma}
\begin{proof}
  All last common ancestors ($\lca$) are taken w.r.t.\ $T$.  By assumption,
  $w\coloneqq \lca(A\cup B)$ is an uncolored inner vertex and thus also an
  uncolored inner vertex of $T[H_i^*]$ for some component $H^*_i$ of $H^*$
  and thus $\lca(x,y)=w$ for all $x\in A$ and $y\in B$.  Moreover, since
  there is no colored vertex on the path $P_{w,\lca(B)}$ with exception of
  $b\coloneqq\lca(B)$, we have $b\in L[H^*_i]$.  Let $e_b=\parent(b) b$ and
  denote by $H_b$ the set of all edges of $H^*$ except the edges along the
  path $P_{u_i,b}$ from the unique child $u_i$ of the root $r_i$ to $b$.
  By Lemma~\ref{lem:componentwise} and the arguments in the proof of
  Lemma~\ref{lem:nonessential}, $H_b$ is a separating set for
  $(T,\mathcal{P})$.  Since $r_i$ only has a single child and
  $w\coloneqq \lca(A\cup B)$ has at least two, we observe that
  $w \preceq u_i$, and thus, $w$ lies along both the path from $u_i$ to
  $b$.  Therefore, $P_{w,b}$ is a subpath of $P_{u_i,w}$ and thus also does
  not contain an edge of $H_b$.  Moreover, for any $y\in B$,
  Lemma~\ref{lem:H-edge-distinct-colors} and $\varpi(b)=B=\varpi(y)$ imply
  that the path $P_{b,y}$ also does not contain an edge in $H_b$.  Since in
  addition, for any $a\in A$ and $b\in B$, the path $P_{\lca(x,y)=w, y}$ is
  composed of the paths $P_{w,b}$ and $P_{b,y}$, we obtain
  $(A,B)\notin \qfitch(T,H_b)$ by Cor.~\ref{cor:edges}.  On the other hand,
  $e_b\in H^*$ is an edge on the path $P_{\lca(x,y)=w, y}$ for all $a\in A$
  and $b\in B$, which, by Cor.~\ref{cor:edges}, implies
  $(A,B)\in \qfitch(T,H^*)$.  In summary, therefore, $(A,B)$ is ambiguous.
\end{proof}

Lemma~\ref{lem:ambiguous} together with the observation that
$(A,B)\notin \qfitch(T,H)$ implies $(B,A)\in \qfitch(T,H)$ for any two
distinct $A,B\in\mathcal{P}$ yields
\begin{corollary}
  \label{cor:ambiguous}
  Let $T$ and $\mathcal{P}$ be compatible and $A,B\in\mathcal{P}$ be
  distinct. If there is a component $H_i^*$ of $H^*$ with $\lca_T(A)$ and
  $\lca_T(B)$ being distinct leaves of $T[H_i^*]$, then there exist choices
  $H_1$, $H_2$, and $H_3$ of separating edge sets such that
  \begin{itemize}
    \item[(1)] $(A,B)\in \qfitch(T,H_1)$ and  $(B,A)\in \qfitch(T,H_1)$,
    \item[(2)] $(A,B)\in \qfitch(T,H_2)$ and $(B,A)\notin \qfitch(T,H_2)$,
      and
    \item[(3)] $(A,B)\notin \qfitch(T,H_3)$ and $(B,A)\in \qfitch(T,H_3)$.
  \end{itemize}
\end{corollary}
The partial results above show that the choices of directions of arc(s)
between $A,B\in\mathcal{P}$ in the quotient graph $\qfitch$ depend in the
$\preceq_T$-order of $\lca(A)$ and $\lca(B)$ and the positioning of colored
vertices along the unique path connecting $\lca(A)$ and $\lca(B)$ in $T$.

\begin{lemma}
  \label{lem:ambiguous2}
  Let $T$ and $\mathcal{P}$ be compatible and $A,B\in\mathcal{P}$ be
  distinct, and suppose there is no colored vertex $v\in V^0(T)$ such that
  $\lca(B)\prec v\preceq \lca(A\cup B)$.  Then $(A,B)$ is forbidden if and
  only if $\lca(A)\prec\lca(B)$, and otherwise $(A,B)$ is ambiguous.
\end{lemma}
\begin{proof}
  Assume that there is no colored vertex $v\in V^0(T)$ such that
  $\lca(B)\prec v\preceq \lca(A\cup B)$.  If $\lca(A)\prec\lca(B)$,
  Prop.~\ref{prop:unambiguous}(ii) implies that
  $(A,B)\notin E(\qfitch(T,H))$ for all choices of $H$.  For the converse,
  suppose that $\lca(A)\not\prec\lca(B)$.  The case $\lca(A)=\lca(B)$ is
  not possible since $A\ne B$ and $T$ and $\mathcal{P}$ are compatible.
  Moreover, $\lca(B)\prec\lca(A)$ is not possible since otherwise the
  colored vertex $\lca(A)$ satisfies $\lca(B)\prec\lca(A)=\lca(A\cup B)$;
  contradicting the assumption.  Therefore, it remains to consider the case
  that $\lca(A)$ and $\lca(B)$ are $\preceq$-incomparable.  By assumption,
  all vertices on the path $P$ except $\lca(B)$ are uncolored.  Therefore,
  we can apply Lemma~\ref{lem:ambiguous} to conclude that $(A,B)$ is
  ambiguous.
\end{proof}

Combining Prop.~\ref{prop:unambiguous} and Lemma~\ref{lem:ambiguous2}
yields the following characterization of edges that are unambiguously
present or absent, and edges whose presence or absence depends on the
choice of $H$.
\begin{theorem}
  \label{thm:direction}
  Let $T$ and $\mathcal{P}$ be compatible and $A,B\in\mathcal{P}$ be
  distinct. Then the following statements hold:
  \begin{description}
  \item[(1)] $(A,B)$ is essential if and only if there is a colored vertex
    $v\in V^0(T)$ such that $\lca(B)\prec v\preceq \lca(A\cup B)$.
  \item[(2)] $(A,B)$ is forbidden if and only if $\lca(A)\prec\lca(B)$.
  \item[(3)] $(A,B)$ is ambiguous if and only if $\lca(A)$ and $\lca(B)$
    are $\preceq$-incomparable and there is no colored vertex $v\in V^0(T)$
    such that $\lca(B)\prec v\preceq \lca(A\cup B)$.
  \end{description}
\end{theorem}
\begin{proof}
  (1) If there is a colored vertex $v\in V^0(T)$ such that
  $\lca(B)\prec v\preceq \lca(A\cup B)$, then $(A,B)$ is essential by
  Prop.~\ref{prop:unambiguous}(i).  If there is no colored vertex
  $v\in V^0(T)$ such that $\lca(B)\prec v\preceq \lca(A\cup B)$, then, by
  Lemma~\ref{lem:ambiguous2}, $(A,B)$ is forbidden or ambiguous and thus
  not essential.

  (2) If $\lca(A)\prec\lca(B)$, then $\lca(B)= \lca(A\cup B)$.  Hence,
  there is no colored vertex $v\in V^0(T)$ such that
  $\lca(B)\prec v\preceq \lca(A\cup B)$.  By Lemma~\ref{lem:ambiguous2},
  this yields that $(A,B)$ is forbidden.  For the converse, recall that by
  \cite[Thm.~4.5]{Hellmuth:21q}, $\lca(A)=\lca(B)$ is not possible.  If
  $\lca(B)\prec\lca(A)$, then $(A,B)$ is essential by
  Cor.~\ref{cor:unambiguous}.  If $\lca(A)$ and $\lca(B)$ are
  $\preceq$-incomparable, then Prop.~\ref{prop:unambiguous}(i) and
  Lemma~\ref{lem:ambiguous2} imply that $(A,B)$ must be essential or
  ambiguous, respectively.  In summary, $\lca(A)\not\prec\lca(B)$ implies
  that $(A,B)$ is not forbidden.

  (3) If $\lca(A)$ and $\lca(B)$ are $\preceq$-incomparable and there is no
  colored vertex $v\in V^0(T)$ such that
  $\lca(B)\prec v\preceq \lca(A\cup B)$, then $(A,B)$ is ambiguous by
  Lemma~\ref{lem:ambiguous2}.  If $\lca(A)$ and $\lca(B)$ are comparable,
  then Cor.~\ref{cor:unambiguous} implies that $(A,B)$ is essential or
  forbidden.  If there is a colored vertex $v\in V^0(T)$ such that
  $\lca(B)\prec v\preceq \lca(A\cup B)$, then $(A,B)$ is essential by
  statement~(1).  By contraposition, therefore, the \emph{only-if}-part of
  statement~(3) is also true.
\end{proof}

\begin{corollary}
  \label{cor:algo}
  Let $T$ and $\mathcal{P}$ be compatible.
  After an $O(|L|)$ preprocessing step, we can query, for two distinct
  $A,B\in \mathcal{P}$, in constant time whether $(A,B)$ is essential,
  forbidden, or ambiguous for $(T,\mathcal{P})$.
  In particular, we can make these assignments for all distinct $A,B\in
  \mathcal{P}$ in $O(|L| + |\mathcal{P}|^2)$ total time.
\end{corollary}
\begin{proof}
  We compute $\varpi$ and $\lca(A)$ for all $\mathcal{P}$ in $O(|L|)$
  according to Lemma~\ref{lem:varpi-compute}.  In particular, this
  procedure already includes the construction an LCA data structure that
  enables constant-time queries for the last common ancestor for any pair
  $v,v'\in V(T)$. We continue by computing, for each $v\in V(T)$, the
  unique $\preceq$-minimal colored vertex $u$ that is a strict ancestor of
  $v$, i.e., that satisfies $v\prec u$, if such a vertex exists.  We
  achieve this by filling a map \texttt{lcsa} (for ``lowest colored strict
  ancestor'') in a top-down traversal of $T$. More precisely, we initialize
  $\texttt{lcsa}(\rho)=\emptyset$ for the root $\rho$. Then, for each
  vertex $v\in V(T)\setminus\{\rho\}$, we set
  $\texttt{lcsa}(v)=\parent_T(v)$ if $\parent_T(v)$ is a colored vertex;
  and $\texttt{lcsa}(v)=\texttt{lcsa}(\parent_T(v))$ otherwise.  We are now
  able to query in constant time, whether, for two vertices $u,v\in V(T)$
  with $v\prec u$, there is a colored vertex $w$ such that
  $v\prec w \preceq u$ by just evaluating whether
  $\lca(\texttt{lcsa}(v),u)=u$.  Checking whether or not
  $\lca(A\cup B)=\lca(\lca(A),\lca(B))$ equals $\lca(B)$ also determines
  whether or not $\lca(A)\prec\lca(B)$ since we know that
  $\lca(A)\ne\lca(B)$ for distinct $A,B\in\mathcal{P}$.  Similarly,
  $\lca(A\cup B) \notin\{\lca(A), \lca(B)\}$ implies that $\lca(A)$ and
  $\lca(B)$ are $\preceq$-incomparable.  In summary, the conditions of
  Thm.~\ref{thm:direction} can therefore be evaluated in constant time for
  each of the $O(|\mathcal{P}|^2)$ pairs $(A,B)$.
\end{proof}

It is important to note that the choice of presence/absence for ambiguous
edges is not arbitrary. First, in the quotient graph $\qfitch(T,H)$ at
least one of $(A,B)$ and $(B,A)$ is always present. Second, $\qfitch(T,H)$
is itself a Fitch graph, and thus characterized by forbidden induced
subgraphs on three vertices \cite{Geiss:18a,Hellmuth:19}. The condition
that an edge must be present in at least one direction reduced the
possibilities to only four allowed configurations on three vertices, namely
the triangles A1, A4, A5, and A6 in Fig.~2 of \cite{Geiss:18a}.

\section{Edges of Fitch Quotient Graphs for Refinements of Trees}
\label{sect:refinements}

Let us now turn to the case that $T$ and $\mathcal{P}$ are 
refinement-compatible but not necessarily compatible. We first collect
some basic results concerning ancestry in trees that is preserved upon
refinement.
\begin{lemma}
  \label{lem:prec-preservation}
  Let $T'$ be a refinement of a tree $T$ with leaf set $L$ and $uv\in E(T)$.
  Then the following statements hold for non-empty $X\subseteq L$.
  \begin{description}
  \item[(i)] If $\lca_{T}(X)\preceq_T v$, then
    $\lca_{T'}(X)\preceq_{T'} v'$ for the vertex $v'\in V(T')$ satisfying
    $L(T(v))=L(T'(v'))$.
  \item[(ii)] If $u\preceq_{T} \lca_{T}(X)$ and
    $L(T(v))\cap X\ne\emptyset$, then $v' \prec_{T'} \lca_{T'}(X)$ for the
    vertex $v'\in V(T')$ satisfying $L(T(v))=L(T'(v'))$.
  \item[(iii)] If $\lca_{T}(X)$ and $\lca_{T}(\tilde X)$ are
    $\preceq_{T}$-incomparable for some $\tilde X\subseteq L$, then
    $\lca_{T'}(X)$ and $\lca_{T'}(\tilde X)$ are
    $\preceq_{T'}$-incomparable.
  \end{description}
\end{lemma}
\begin{proof}
  Note first that, since $T'$ is a refinement of $T$, we have
  $\mathcal{H}(T)\subseteq \mathcal{H}(T')$, and thus the vertex
  $v'\in V(T')$ satisfying $L(T(v))=L(T'(v'))$ exists for every
  $v\in V(T)$.  First suppose $\lca_{T}(X)\preceq_T v$. Therefore, we have
  $X \subseteq L(T(v))=L(T'(v'))$. Together with the fact that $X$ is
  non-empty, this implies that $\lca_{T'}(X)\preceq_{T'} v'$, and thus
  statement~(i).  Suppose $u\preceq_{T} \lca_{T}(X)$ and
  $L(T(v))\cap X\ne\emptyset$. Recall that, by convention, $uv\in E(T)$
  implies that $v\prec_T u$.  Hence, $v\prec_T u\preceq_{T} \lca_{T}(X)$
  and, therefore, $X \not\subseteq L(T(v))=L(T'(v'))$.  Since by assumption
  there is an $x\in X$ such that $x \in L(T(v))=L(T'(v'))$, $v'$ and
  $\lca_{T'}(X)$ are both ancestors of $x$ and thus
  $\preceq_{T'}$-comparable. Together with $X \not\subseteq L(T'(v'))$,
  this yields $v' \prec_{T'} \lca_{T'}(X)$ and thus statement~(ii) is true.
  Finally, suppose $\lca_{T}(X)$ and $\lca_{T}(\tilde X)$ are
  $\preceq_{T}$-incomparable for some $\tilde X\subseteq L$, set
  $u=\lca_{T}(X)$ and $\tilde u=\lca_{T}(\tilde X)$, and let
  $u', \tilde{u}'\in V(T')$ be the vertices satisfying $L(T(u))=L(T'(u'))$
  and $L(T(\tilde u))=L(T'(\tilde u'))$.  Since $u$ and $\tilde u$ are
  $\preceq_{T}$-incomparable, we have
  $\emptyset=L(T(u))\cap L(T(\tilde u))=L(T(u'))\cap L(T(\tilde u'))$, and
  thus $u'$ and $\tilde{u}'$ are $\preceq_{T'}$-incomparable. Moreover,
  $X\subseteq L(T(u))=L(T'(u'))$ and
  $\tilde X\subseteq L(T(\tilde u))=L(T'(\tilde u'))$ implies
  $\lca_{T'}(X)\prec_{T'}u'$ and $\lca_{T'}(\tilde X)\prec_{T'} \tilde
  u'$. This together with $u'$ and $\tilde{u}'$ being
  $\preceq_{T'}$-incomparable implies statement~(iii).
\end{proof}
We next derive sufficient conditions for pairs $(A,B)$ that are r-essential
or r-forbidden, respectively.
\begin{proposition}
  \label{prop:unambiguous-refinement}
  Let $T$ and $\mathcal{P}$ be r-compatible. Then for all
  $A,B\in \mathcal{P}$ it holds
  \begin{description}
    \item[(i)] If the path $P_{\lca_{T}(A\cup B), \lca_{T}(B)}$ contains a
    colored edge, then $(A,B)$ is r-essential.
    \item[(ii)] If $\lca_{T}(A) \prec_T \lca_{T}(B)$ and
    $\gamma(\lca_{T}(B)v)=B$ for the vertex $v\in \child_{T}(\lca_{T}(B))$
    satisfying $\lca_{T}(A)\preceq_T v$, then
    $(A,B)$ is r-forbidden.
    \item[(iii)] If the path $P_{u, \lca_{T}(A)}$ with $u\coloneqq
    \lca_{T}(A\cup B)\ne\rho_T$ contains a colored edge $e$ with
    $\gamma(e)=\gamma(\parent_T(u) u)=C\in\mathcal{P}$, then
    $(A,B)$ is r-essential.
  \end{description}
\end{proposition}
\begin{proof}
  All three statements are proven by considering an arbitrary refinement
  $T'$ of $T$ and a set $H\subseteq E(T')$ satisfying
  $\partTH(T',H)=\mathcal{P}$.  In particular, the vertex coloring
  $\varpi \colon V(T')\to \mathcal{P} \cup \{\emptyset\}$ is well-defined
  for any such $T'$ and $\mathcal{P}$.

  (i) Suppose that there is an edge
  $uv\in E(P_{\lca_{T}(A\cup B), \lca_{T}(B)})$ with $\gamma(uv)=C$ for
  some $C\in \mathcal{P}$.  In particular, this implies
  $\lca_{T}(B)\prec_T\lca_{T}(A\cup B)$.  We have to show that
  $(A,B)\in \qfitch(T',H)$.
  Lemma~\ref{lem:prec-preservation}(i) together with
  $\lca_{T}(B)\preceq_T v$ implies $\lca_{T'}(B)\preceq_{T'} v'$ where $v'$
  is the vertex in $T'$ satisfying $L(T'(v'))=L(T(v))$.  By assumption,
  $u\preceq_{T} \lca_{T}(A\cup B)$ and
  $\emptyset\ne B\subseteq L(T(v))\cap (A\cup B)$. Together with
  Lemma~\ref{lem:prec-preservation}(ii), this implies
  $v' \prec_{T'} \lca_{T'}(A\cup B)$.  Since $\gamma(uv)=C$, we have
  $L(T'(v'))\cap C=L(T(v))\cap C\ne \emptyset$ and also
  $L(T'(v'))\setminus C=L(T(v))\setminus C\ne \emptyset$.  Therefore, $v'$
  lies on the path that connects two distinct vertices from $C$, and thus
  $\varpi(v')=C$.  Moreover, since $B\subseteq L(T(v))=L(T'(v'))$ and
  $L(T'(v'))\setminus C\ne \emptyset$, we have $B\ne C$. Together with
  $\varpi(\lca_{T'}(B))=B$ and $\varpi(v')=C$, this implies
  $\lca_{T'}(B)\ne v'$, and thus $\lca_{T'}(B)\prec_{T'} v'$.  In summary,
  there is a colored vertex $v'\in V^0(T')$ such that
  $\lca_{T'}(B)\prec_{T'} v'\preceq_{T'} \lca_{T'}(A\cup
  B)=\lca_{T'}(\lca_{T'}(A), \lca_{T'}(B))$.  Hence, we can apply
  Thm.~\ref{thm:direction}(1) to conclude that the statement is true.

  (ii) Suppose $\lca_{T}(A) \prec_T \lca_{T}(B)$ and
  $\gamma(\lca_{T}(B)v)=B$ for the vertex $v\in \child_{T}(\lca_{T}(B))$
  satisfying $\lca_{T}(A)\preceq_T v$.  We have to show that
  $(A,B)\notin \qfitch(T',H)$.  Lemma~\ref{lem:prec-preservation}(i)
  together with $\lca_{T}(A)\preceq_T v$ implies
  $\lca_{T'}(A)\preceq_{T'} v'$ where $v'$ is the vertex in $T'$ satisfying
  $L(T'(v'))=L(T(v))$.  Since $\gamma(\lca_{T}(B)v)=B$, we have
  $L(T(v))\cap B\ne\emptyset$ which together with
  Lemma~\ref{lem:prec-preservation}(ii) implies
  $v' \prec_{T'} \lca_{T'}(B)$.  In summary, we have
  $\lca_{T'}(A)\preceq_{T'} v' \prec_{T'} \lca_{T'}(B)$.  Now apply
  Prop.~\ref{prop:unambiguous}(ii).

  (iii) By assumption, there is an edge $vw\in E(P_{u,\lca_{T}(A)})$, where
  $u\coloneqq \lca_{T}(A\cup B)$, of color
  $\gamma(vw)=C = \gamma(\parent_T(u) u)$.  We observe that $A\ne C$, and
  $L(T(w))\cap C\ne\emptyset$.  Lemma~\ref{lem:prec-preservation}(ii),
  $v\preceq_{T} u=\lca_{T}(A\cup B)$, and $L(T(w))\cap C\ne\emptyset$ imply
  $w'\prec_{T'} \lca_{T'}(A\cup B)\eqqcolon u'$ for the vertex
  $w'\in V(T')$ satisfying $L(T'(w'))=L(T(w))$.  From
  $\gamma(\parent_T(u) u)=C$, we conclude $C \setminus L(T(u))\ne\emptyset$
  and since $\lca_{T}(B)\preceq_{T} u$ also $C\ne B$.  Since moreover
  $L(T'(u')) \subseteq L(T(u))$, we also have
  $C \setminus L(T'(u'))\ne\emptyset$.  Therefore and since there is a
  vertex $c\in C$ with $c \preceq_{T'} w'\prec_{T'} u'$, we must have
  $c \preceq_{T'} w'\prec_{T'} u'\prec_{T'} \lca_{T}(C)$.  In particular,
  therefore, vertex $u'$ lies on the path connecting two vertices of $C$,
  and thus it is a vertex of color $\varpi(u')=C$ on the path
  $P_{u', \lca_{T'}(B)}$. We have $\varpi(u')=C\ne B=\varpi(\lca_{T'}(B))$
  and thus $\lca_{T'}(B)\prec_{T'} u'$. Now apply
  Thm.~\ref{thm:direction}(1).
\end{proof}
We note that the conditions in Prop.~\ref{prop:unambiguous-refinement}(ii) are
a special case of those in Prop.~\ref{prop:unambiguous-refinement}(i), since
the edge $\lca_{T}(A)v$ is a colored edge on the path $P_{\lca_{T}(A\cup B),
\lca_{T}(B)}$, and thus we obtain
\begin{corollary}
  \label{cor:unambiguous-refinement}
  Let $T$ and $\mathcal{P}$ be r-compatible.
  If, for $A,B\in \mathcal{P}$, it holds $\lca_{T}(A) \prec_T \lca_{T}(B)$ and
  $\gamma(\lca_{T}(B)v)=B$ for the vertex $v\in \child_{T}(\lca(B))$
  satisfying $\lca_{T}(A)\preceq_T v$, then
  $(B,A)$ is r-essential and $(A,B)$ is r-forbidden.
\end{corollary}
Similar to the previous section, we will show that it is not possible to
unambiguously assign a direction for all pairs $(A,B)$ that are not covered
by Prop.~\ref{prop:unambiguous-refinement}.  Even though we do not want to
construct the set of all compatible refinement explicitly, it will be
helpful to study the properties of two special representatives.  A
compatible refinement $T'$ for an r-compatible pair $(T,\mathcal{P})$ can
be constructed in linear time \cite[Thm.~7.5]{Hellmuth:21q}. To this end,
\citet{Hellmuth:21q} considered the set
\begin{equation}
  \mathfrak{Y}(T,\mathcal{P}) \coloneqq \left\{
    A\in\mathcal{P}\mid \exists B\in\mathcal{P}\colon B\ne A,\,
    B\cap\cl(A)\ne\emptyset \text{ and } \cl_{T}(A)\subseteq
    \cl_{T}(B)
  \right\}\,,
\end{equation}
which contains the sets $A\in\mathcal{P}$ for which the vertex
$u\coloneqq\lca_{T}(A)\in V(T)$ is ``not resolved enough''.  The conditions
$B\cap\cl_{T}(A)\ne\emptyset$ and $\cl_{T}(A)\subseteq \cl_{T}(B)$ are
equivalent to $u$ having a child $v$ such that $\gamma(uv)=B$.
\begin{figure}
  \begin{center}
    \includegraphics[width=0.8\textwidth]{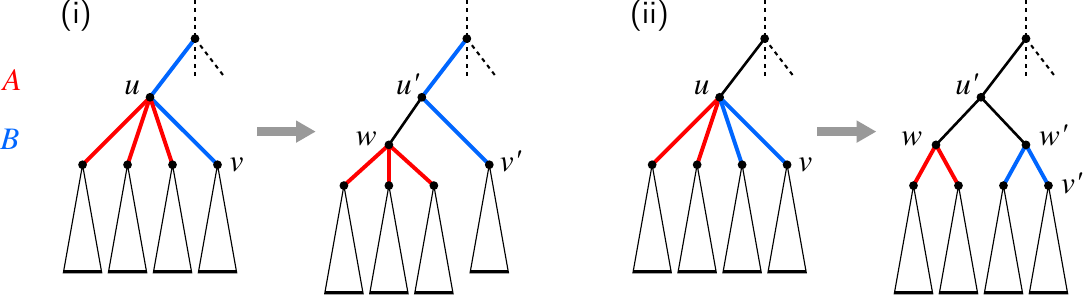}
  \end{center}
  \caption{Necessary refinement steps for a set
    $A\in \mathfrak{Y}(T,\mathcal{P})$. In both examples, there is a
    vertex $v\in\child_{T}(u\coloneqq\lca_{T}(A))$ such that $\gamma(uv)=B$
    for some $B\in\mathcal{P}\setminus \{A\}$.  The two cases (i)
    $u\prec_{T} \lca_{T}(B)$ and (ii) $u= \lca_{T}(B)$ are possible. The
    latter case implies that also $B\in \mathfrak{Y}(T,\mathcal{P})$.}
  \label{fig:Yset}
\end{figure}
To see this, suppose first $B\cap\cl_T(A)\ne\emptyset$ and
$\cl_{T}(A)\subseteq \cl_{T}(B)$.  Since $A\ne B$ and
$B\cap L(T(u)) =B\cap\cl_T(A)\ne\emptyset$, we can conclude that $L(T(u))$
contains at least one vertex of both $A$ and $B$. In particular, $u$ is an
inner vertex and has a child $v$ such that $B\cap
L(T(v))\ne\emptyset$. However, $\cl_{T}(A)\subseteq\cl_{T}(B)$ is
equivalent to $u\preceq_T \lca_{T}(B)$ and thus
$B\setminus L(T(v))\ne\emptyset$. Therefore, the edge $uv$ lies on the path
connecting two vertices from $B$, and thus $\gamma(uv)=B$.  Conversely, if
$\gamma(\lca_{T}(A)v)=B$, then $B\cap L(T(v))\ne\emptyset$ and
$B\setminus L(T(v))\ne\emptyset$ implying $B\cap\cl_T(A)\ne\emptyset$ and
$\cl_{T}(A)\subseteq \cl_{T}(B)$, respectively, see Fig.~\ref{fig:Yset} for
a graphical depiction of the situation.

We thus obtain
\begin{equation}
  \mathfrak{Y}(T,\mathcal{P}) = \left\{
  A\in\mathcal{P}\mid \exists v\in \child_{T}(\lca_{T}(A)) \colon
  \gamma(\lca_{T}(A)v)=B \in \mathcal{P}\setminus\{A\})
  \right\}.
\end{equation}
By \cite[Lemma~5.6]{Hellmuth:21q}, the tree $T^*$ satisfying
\begin{equation}
  \mathcal{H}(T^*) = \mathcal{H}(T) \cup \bigcup_{A\in
  \mathfrak{Y}(T,\mathcal{P})} \{
     \bigcup_{\quad\substack{ v\in\child_{T}(\lca_{T}(A)),\\
                           \gamma(\lca_{T}(A)v)=A}
                         } L(T(v)) \quad \}
\end{equation}
is a refinement of $T$ that is compatible with $\mathcal{P}$.
Algorithmically, $T^*$ is obtained by introducing, for each
$A\in\mathfrak{Y}(T,\mathcal{P})$, a new inner vertex $w_A$ that replaces
all edges $\lca_{T}(A)v$ satisfying $\gamma(\lca_{T}(A)v)=A$ by edges
$w_A v$ and re-connects $w_A$ as a new child of $\lca_{T}(A)$ (cf.\ proof
of \cite[Thm.~7.5]{Hellmuth:21q}).

The tree $T^*$ is still ``not resolved enough'' to remove all possible
constraints on the direction of edges.  More precisely, the fact that
$(A,B)$ is r-ambiguous does not imply that $(A,B)$ is ambiguous w.r.t.\
$T^*$.  A counterexample is shown in Fig.~\ref{fig:qfitch-example}(ii).
Here we have $\mathfrak{Y}(T_{3},\mathcal{P})=\emptyset$ and thus
$T^*=T_3$. By Cor.~\ref{cor:unambiguous},
$\lca_{T^*}(B)\prec_{T^*} \lca_{T^*}(A)$ implies $(A,B)\in \qfitch(T^*,H)$
and $(B,A)\notin \qfitch(T^*,H)$ for every valid choice of
$H\subseteq E(T^*)$. However, $T^*$ can be further refined to the tree
$T_4$ in Fig.~\ref{fig:qfitch-example}(iii), which does not constrain the
direction of the edges between $A$ and $B$ due to
Lemma~\ref{lem:ambiguous2}.

In the following, we will say that the leaf set of a subtree
$L(T(v))\in\mathcal{H}(T)\setminus\{L(T)\}$ of $T$ is \emph{colored (by
  $A\in \mathcal{P}$)} if $\gamma(\parent_T(v)v)=A$.  If there is no such
set $A\in \mathcal{P}$, then we say that $L(T(v))$ is uncolored.
\begin{fact}
  Let $T$ and $\mathcal{P}$ be r-compatible. Then $L(T(v))$ is colored by
  $A\in\mathcal{P}$ if and only if $A \cap L(T(v))\ne \emptyset$ and
  $A \setminus L(T(v))\ne \emptyset$.
\end{fact}

\begin{definition}
  \label{def:basic-refinement}
  Let $T$ be a tree. A \emph{basic refinement step (on $X$)} is an
  operation that takes, for some $u\in V^0(T)$, a subset
  $X \subsetneq \child_{T}(u)$ with $|X|>1$ and
  \begin{description}
    \item[(1)] introduces a new inner vertex $w$,
    \item[(2)] replaces all edges $uv$ by edges $wv$ for all $v\in X$, and
    \item[(3)] re-connects $w$ as a new child of $u$.
  \end{description}
\end{definition}
In particular, observe that $|\mathcal{H}(T')\setminus \mathcal{H}(T)|=1$
if $T'$ is a refinement of $T$ obtained by a basic refinement step, and
thus, the hierarchy $\mathcal{H}(T')$ differs from $\mathcal{H}(T)$ by
exactly one additional set.  Moreover, if $T$ and $\mathcal{P}$ are
r-compatible, then the tree $T^*$ is obtained by a series basic refinement
steps.
\begin{definition}
  \label{def:TP}
  Let $T$ and $\mathcal{P}$ be r-compatible. A basic refinement step,
  resulting in a tree $T'$, is \emph{uncolored} if the set $L(T'(w))$
  inserted into $\mathcal{H}(T)$ is uncolored, i.e., if
  $\gamma(\parent_{T'}(w)w)=\emptyset$.  An \emph{uncolored refinement step
    (URS)-tree} for $(T,\mathcal{P})$ is a tree $\TP$ that (i) is obtained
  from $T^*$ by a sequence of uncolored refinement steps and (ii) does not
  admit an uncolored refinement step.
\end{definition}
Since the tree $T^*$ can be constructed for every r-compatible pair
$(T,\mathcal{P})$, it is also always possible to construct a URS-tree $\TP$
according to Def.~\ref{def:TP}.

\begin{lemma}
  \label{lem:uncolored-refinement}
  Let $T$ and $\mathcal{P}$ be r-compatible and $A\in \mathcal{P}$.
  Suppose that $u\coloneqq \lca_{T}(A)$ is an inner vertex and let $V'$ be
  set of children $v\in\child_{T}(u)$ for which $L(T(v))$ is colored by
  $A$.  If $V'\subsetneq \child_{T}(u)$, then the basic refinement step on
  $V'$ is uncolored.
\end{lemma}
\begin{proof}
  Since $T$ and $\mathcal{P}$ be r-compatible, the coloring of the elements
  in $\mathcal{H}$ is well-defined. We observe first that since
  $u=\lca_{T}(A)$ is an inner vertex, we have $|V'|\ge 2$, and thus, we
  consider a valid basic refinement step. More precisely, a new inner
  vertex $w$ is introduced and, for all $v\in V'$, the edge $\lca_{T}(A)v$
  is replaced the edge $wv$, and $w$ is re-connected as a new child of
  $\lca_{T}(A)$.  Hence, the refinement step introduces a set $W$ into the
  corresponding hierarchy that is the union of the sets $L(T(v))$ of the
  vertices $v\in V'$.  Moreover, we have $A\subseteq W$ by construction and
  thus $W$ is not colored by $A$.  Now assume that $W$ is colored by $B$
  for some $B\in \mathcal{P}\setminus \{A\}$, i.e., $B \cap W\ne \emptyset$
  and $B \setminus W\ne \emptyset$.  By construction, there must be some
  $v\in V'$ such that there is a $b\in B$ with $b\in L(T(v))\subsetneq
  W$. This and $B \setminus W\ne \emptyset$ implies
  $B \cap L(T(v))\ne \emptyset$ and $B \setminus L(T(v))\ne \emptyset$.
  Thus $L(T(v))$ is colored by $B$; a contradiction to $L(T(v))$ being
  colored by $A$.  Therefore, $W$ must be uncolored.
\end{proof}

\begin{lemma}
  \label{lem:TP-properties}
  Let $T$ and $\mathcal{P}$ be r-compatible and $\TP$ a corresponding
  URS-tree.  Then $\TP$ and $\mathcal{P}$ are compatible and every set in
  $\mathcal{H}(\TP) \setminus \mathcal{H}(T)$ is uncolored.
\end{lemma}
\begin{proof}
  By construction, $\TP$ is a refinement of $T^*$ and thus
  $\mathcal{H}(T^*)\subseteq \mathcal{H}(\TP)$.  Since moreover $T^*$ and
  $\mathcal{P}$ are compatible and refinements preserve compatibility
  \cite[Prop.~5.1]{Hellmuth:21q}, $\TP$ is also compatible with
  $\mathcal{P}$, and thus also r-compatible.  We have
  $\mathcal{H}(T)\subseteq \mathcal{H}(T^*)\subseteq \mathcal{H}(\TP)$.
  The tree $T^*$ is obtained from $T$ by a series of basic refinement steps
  as described above.  To show that these basic refinement steps are
  uncolored, we consider the first step, which refines the last common
  ancestor of some set
  $A\in\mathfrak{Y}(T,\mathcal{P})\subseteq \mathcal{P}$. Let $V'$ be set
  of children $v\in\child_{T}(\lca_{T}(A))$ with $\gamma(\lca_{T}(A)v)=A$,
  i.e., for which $L(T(v))$ is colored by $A$.  By construction of
  $\mathfrak{Y}(T,\mathcal{P})$, we have $V'\neq
  \child_{T}(\lca_{T}(A))$. By Lemma~\ref{lem:uncolored-refinement}, this
  basic refinement step on $V'$ is uncolored.  Since the order of the basic
  refinement steps leading to $T^*$ is arbitrary, every set in
  $\mathcal{H}(T^*)\setminus \mathcal{H}(T)$ must be uncolored.  In
  addition, every set in $\mathcal{H}(\TP)\setminus \mathcal{H}(T^*)$ is
  uncolored by construction.  In summary, therefore, every set in
  $\mathcal{H}(\TP) \setminus \mathcal{H}(T)$ is uncolored.
\end{proof}

\begin{lemma}
  \label{lem:ambiguous-refinement}
  Let $T$ and $\mathcal{P}$ be r-compatible, $A,B\in\mathcal{P}$ be
  distinct, and set $u\coloneqq \lca_{T}(A\cup B)$.  Suppose there is no
  colored edge $e\in E(P_{u, \lca_{T}(B)})$ and the path
  $P_{u, \lca_{T}(A)}$ contains no colored edge $e$ with
  $\gamma(e)=\gamma(\parent_T(u) u)=C\in\mathcal{P}$.  Then $(A,B)$ is
  r-forbidden if and only if $\lca_T(A)\prec\lca_T(B)$ and
  $\gamma(\lca_{T}(B)v)=B$ for the vertex $v\in \child_{T}(\lca_{T}(B))$
  satisfying $\lca_{T}(A)\preceq_T v$. Otherwise $(A,B)$ is r-ambiguous.
\end{lemma}
\begin{proof}
  Suppose first that $\lca_T(A)\prec\lca_T(B)$ and $\gamma(\lca_{T}(B)v)=B$
  for the vertex $v\in \child_{T}(\lca_{T}(B))$ satisfying
  $\lca_{T}(A)\preceq_T v$.  By
  Prop.~\ref{prop:unambiguous-refinement}(ii), $(A,B)$ is r-forbidden.

  For the converse, suppose that $\lca_T(A)\not\prec\lca_T(B)$ or
  $\gamma(\lca_{T}(B)v)\ne B$.  We consider a corresponding URS-tree $\TP$,
  which exists since $T$ and $\mathcal{P}$ are r-compatible.  We write
  $\widehat{u}\eqqcolon \lca_{\TP}(A\cup B)$ and observe that $\widehat{u}$
  is an inner vertex.  By Lemma~\ref{lem:TP-properties}, $\TP$ and
  $\mathcal{P}$ are compatible.  In particular, the vertex coloring
  $\varpi$ is defined for $\TP$ and $\mathcal{P}$.  The edge coloring, on
  the other hand, is defined for both $(T, \mathcal{P})$ and
  $(\TP, \mathcal{P})$. We will therefore write $\gamma_T$ and
  $\gamma_{\TP}$, respectively.
  \begin{claim}
    The two vertices $\lca_{\TP}(A)$ and $\lca_{\TP}(B)$ are
    $\preceq_{\TP}$-incomparable.
  \end{claim}
  \begin{claim-proof}
    Since $\TP$ and $\mathcal{P}$ are compatible, we have
    $\lca_{\TP}(A)\ne \lca_{\TP}(B)$.  Suppose, for contradiction, that
    $\lca_{\TP}(A) \prec_{\TP} \lca_{\TP}(B)=\widehat{u}$ and let
    $\widehat{w}\in \child_{\TP}(\widehat{u})$ be the vertex satisfying
    $\lca_{\TP}(A) \preceq_{\TP} \widehat{w}$.  We distinguish cases (a)
    $L(\TP(\widehat{w}))$ is colored by $B$, and (b) $L(\TP(\widehat{w}))$
    is not colored by $B$. In any of these two cases, we have
    $A\subseteq L(\TP(\widehat{w}))$ and
    $L(\TP(\widehat{w}))\setminus B\ne\emptyset$.

    In case~(a), $L(\TP(\widehat{w}))$ being colored by $B$ and
    Lemma~\ref{lem:TP-properties} imply that
    $L(\TP(\widehat{w}))\in\mathcal{H}(T)$.  Thus, let $w\in V(T)$ be the
    vertex satisfying $L(T(w))=L(\TP(\widehat{w}))$.  Since
    $L(T(w))=L(\TP(\widehat{w}))$ is colored by $B$, i.e.,
    $B \cap L(T(w))\ne\emptyset$ and $B \setminus L(T(w))\ne\emptyset$, we
    have that $w\prec_T \lca_{T}(B)$.  Moreover,
    $A\subseteq L(\TP(\widehat{w}))=L(T(w))$ yields
    $\lca_{T}(A)\preceq_{T} w$, and thus
    $\lca_{T}(A)\preceq_{T} w \prec_T \lca_{T}(B)$.  Now consider the child
    $v\in \child_{T}(\lca_{T}(B))$ such that $\lca_{T}(A)\preceq_{T} v$.
    Clearly, it holds $w\preceq_{T} v$ which, together with
    $B \cap L(T(w))\ne\emptyset$, implies $B \cap L(T(v))\ne\emptyset$.  We
    also have $B \setminus L(T(v))\ne\emptyset$ since
    $v\prec_T \lca_{T}(B)$.  Hence, $\lca_{T}(A) \prec_T \lca_{T}(B)$ and
    $\gamma_T(\lca_{T}(B)v)= B$; a contradiction to the assumption.

    In case~(b), $L(\TP(\widehat{w}))$ is not colored by $B$. Together with
    $B\setminus L(\TP(\widehat{w}))\ne\emptyset$, this yields
    $B\cap L(\TP(\widehat{w}))\ne\emptyset$.  However, since $\widehat{u}$
    is an inner vertex, the set $V'\subseteq \child_{\TP}(\widehat{u})$ of
    vertices $\widehat{w}'$ satisfying
    $B\cap L(\TP(\widehat{w}'))\ne\emptyset$ must contain at least two
    vertices.  In particular, $\widehat{w}\notin V'$ and thus
    $V'\subsetneq \child_{\TP}(\widehat{u})$.  Now consider the basic
    refinement step on $V'$.  By Lemma~\ref{lem:uncolored-refinement}, this
    refinement step is uncolored; a contradiction to $\TP$ being a URS-tree.

    Now suppose, for contradiction, that
    $\lca_{\TP}(B) \prec_{\TP} \lca_{\TP}(A)=\widehat{u}$ and let
    $\widehat{w}\in \child_{\TP}(\widehat{u})$ be the vertex satisfying
    $\lca_{\TP}(B) \preceq_{\TP} \widehat{w}$.  We distinguish cases (a')
    $L(\TP(\widehat{w}))$ is colored by $A$, and (b) $L(\TP(\widehat{w}))$
    is not colored by $A$.  We can apply similar arguments as in cases~(a)
    and (b) to conclude that the path $P_{\widehat{u},\lca_{T}(B)}$
    contains an edge of color $A$ or that $\TP$ admits an uncolored basic
    refinement step, respectively, both of which contradict the assumptions
    In summary, therefore, $\lca_{\TP}(A)$ and $\lca_{\TP}(B)$ are
    $\preceq_{\TP}$-incomparable.
  \end{claim-proof}

  \begin{claim}
    \label{claim:P}
    The path $P\coloneqq P_{\widehat{u},\lca_{\TP}(B)}$ in $\TP$ does not
    contain a colored edge.
  \end{claim}
  \begin{claim-proof}
    Assume, for contradiction, that the path $P$ contains a colored edge
    $\widehat{p}\widehat{q}$.  By definition, this implies that
    $\varpi(\widehat{p})= \varpi(\widehat{q})=C$ for some
    $C\in\mathcal{P}$.  Since $\lca_{\TP}(B)\preceq_{\TP} \widehat{q}$ and
    $\varpi(\lca_{\TP}(B))=B$, we have $\lca_{\TP}(B)\ne \widehat{q}$, and
    thus $\lca_{\TP}(B)\prec_{\TP} \widehat{q}\prec_{\TP} w$.  From
    $\varpi(\widehat{q})=C$ and
    $\varpi(\parent_{\TP}(\widehat{q}))= \varpi(\widehat{p})=C$, we
    conclude that $C\cap L(\TP(\widehat{q}))\ne\emptyset$ and
    $C\setminus L(\TP(\widehat{q}))\ne\emptyset$, respectively.  Hence,
    $L(\TP(\widehat{q}))$ is colored by $C$, which together with
    Lemma~\ref{lem:TP-properties} implies that
    $L(\TP(\widehat{q}))\in \mathcal{H}(T)$.  Hence, let $q\in V(T)$ be the
    vertex satisfying $L(T(q))=L(\TP(\widehat{q}))$.  Since
    $\widehat{p}\widehat{q}$ is an edge on the path $P$, we have
    $\emptyset\ne B\subseteq L(\TP(\widehat{q}))=L(T(q))$, and thus,
    $\lca_{T}(B)\preceq_{T} q$.  Moreover, since $\lca_{\TP}(A)$ and
    $\lca_{\TP}(B)$ are incomparable and
    $\lca_{\TP}(B)\preceq_{\TP} \widehat{q}\prec_{\TP} \widehat{p}
    \preceq_{\TP} \widehat{u}$, we have
    $A\cap L(\TP(\widehat{q}))=\emptyset$, and thus,
    $A\cap L(T(q))=\emptyset$.  This, together with the fact that $q$ and
    $u=\lca_{T}(A\cup B)$ are both ancestors of $\lca_{T}(B)$ and thus
    $\preceq_{T}$-comparable, implies $q\prec_T u$.  Moreover,
    $C\cap L(T(q))\ne\emptyset$ and $C\setminus L(T(q))\ne\emptyset$
    implies that the edge $\parent_T(q)q$ connects two vertices in $C$,
    i.e., we have $\gamma_T(\parent_T(q)q)=C$.  Clearly, therefore,
    $\gamma_T(\parent_T(q)q)=C$ is a colored edge on the path
    $P_{u,\lca_{T}(B)}$; a contradiction.  Hence, there cannot be a colored
    edge in $P$.
  \end{claim-proof}

  \begin{claim}
    The path $P$ in $\TP$ does not contain a colored vertex
    $\widehat{v}\in V^0(\TP)$ such that
    $\lca_{\TP}(B)\prec_{\TP} \widehat{v}\preceq_{\TP} \widehat{u}$.
  \end{claim}
  \begin{claim-proof}
    Suppose, for contradiction, that $P$ contains a vertex $\widehat{v}$
    such that
    $\lca_{\TP}(B)\prec_{\TP} \widehat{v}\preceq_{\TP} \widehat{u}$ and
    $\varpi(\widehat{v})=C\in\mathcal{P}$.  Since
    $\lca_{\TP}(B)\prec_{\TP} \widehat{v}$, we know that $\widehat{v}$ is
    an inner vertex and $B\ne C$.  Let
    $\widehat{w}\in \child_{\TP}(\widehat{v})$ be the vertex satisfying
    $\lca_{\TP}(B)\preceq_{\TP} \widehat{w}$.  We must have
    $C\cap L(\TP(\widehat{w}))=\emptyset$ since otherwise the edge
    $\widehat{v}\widehat{w}\in E(P)$ lies on the path connecting two
    vertices in $C$ and thus, it is a colored edge;
    contradicting the previous claim.  We distinguish the two
    cases~(a) $\widehat{v}=\lca_{\TP}(C)$ and~(b)
    $\widehat{v}\ne\lca_{\TP}(C)$.

    In case~(a), $\widehat{v}=\lca_{\TP}(C)$ implies that
    $V'\coloneqq \{\widehat{w}'\in \child_{\TP}(\widehat{v}) \mid C\cap
    L(\TP(\widehat{w}'))\ne\emptyset \}$ contains at least two vertices.
    In particular, $\widehat{w}\notin V'$ and thus
    $V'\subsetneq \child_{\TP}(\widehat{v})$.  Hence, all conditions of
    Lemma~\ref{lem:uncolored-refinement} are satisfied.  The basic
    refinement step on $V'$ is therefore uncolored; a contradiction to
    $\TP$ being a URS-tree.

    In case~(b), $\varpi(\widehat{v})=C$ (and thus
    $C\cap L(\TP(\widehat{v}))\ne\emptyset$) together with
    $\widehat{v}\ne\lca_{\TP}(C)$ imply that $L(\TP(\widehat{v}))$ is
    colored by $C$ and, equivalently,
    $\gamma_{\TP}(\parent_{\TP}(\widehat{v})\widehat{v})=C$.  Clearly, this
    is possible only if $\widehat{v}=\widehat{u}$ since $P$ does not contain a
    colored edge.
    Moreover, $L(\TP(\widehat{u}))$ being colored by $C$ and
    Lemma~\ref{lem:TP-properties} imply that
    $L(\TP(\widehat{u}))\in\mathcal{H}(T)$.  In particular, we have
    $\gamma_T(\parent_T(u')u')=C$ for the vertex $u'\in V(T)$ satisfying
    $L(T(u'))=L(\TP(\widehat{u}))$.  Suppose for contradiction that
    $\lca_{T}(A\cup B)=u\ne u'$.  By assumption,
    $A\cup B \in L(\TP(\widehat{u}))=L(T(u'))$ and thus $u\prec_{T} u'$. In
    turn, this implies $L(T(u))\subsetneq L(T(u'))$.  Together with
    $\mathcal{H}(T)\subseteq \mathcal{H}(\TP)$, we conclude that there is a
    vertex $\widehat{u}''\in V(\TP)$ satisfying
    $A \cup B\subseteq L(\TP(\widehat{u}'')) = L(T(u))\subsetneq L(T(u'))=
    L(\TP(\widehat{u}))$; a contradiction to
    $\widehat{u}=\lca_{\TP}(A\cup B)$.  Hence, we must have $u=u'$ and thus
    $\gamma_T(\parent_T(u)u)=C$.  Let
    $\widehat{z}\in \child_{\TP}(\widehat{u})$ be the vertex satisfying
    $\lca_{\TP}(A)\preceq_{\TP} \widehat{z}$.  Suppose, for contradiction,
    that the edge $\widehat{u}\widehat{z}$ is colored.  We must have
    $\gamma_{\TP}(\widehat{u}\widehat{z})=C$ since $\varpi(\widehat{u})=C$
    and $\varpi$ is well-defined for the compatible pair
    $(\TP,\mathcal{P})$.  In particular,
    $\gamma_{\TP}(\widehat{u}\widehat{z})=C$ is equivalent to
    $L(\TP(\widehat{z}))$ being colored by $C$, which together with
    Lemma~\ref{lem:TP-properties} implies
    $L(\TP(\widehat{z}))\in\mathcal{H}(T)$.  Thus, let $z\in V(T)$ be the
    vertex satisfying $L(T(z))=L(\TP(\widehat{z}))$.  From
    $A\subseteq L(\TP(\widehat{z})) \subsetneq L(\TP(\widehat{u}))$ and the
    correspondence of the vertices in $\TP$ and $T$, we obtain
    $\lca_{T}(A)\preceq_{T} z \prec_{T} u$.  Together
    with $C\cap L(T(z))\ne\emptyset$ and $C\setminus L(T(u))\ne\emptyset$,
    this implies that all edges $e$ on the path
    $P_{u,z}\subseteq P_{u,\lca_{T}(A)}$ lie on paths connecting vertices
    from $C$ and are thus colored by $C$. In particular, there is at least
    one such edge $e$ on the path $P_{u,\lca_{T}(A)}$ with
    $\gamma_T(e)=\gamma_T(\parent_T(u)u)=C$; a contradiction.  Hence, the
    edge $\widehat{u}\widehat{z}$ in $\TP$ must be uncolored.  In
    particular, we have $C\cap L(\TP(\widehat{z}))=\emptyset$.  Let
    $\widehat{z}'\in \child_{\TP}(\widehat{u})$ be the vertex satisfying
    $\lca_{\TP}(B)\preceq_{\TP} \widehat{z}'$.  The edge
    $\widehat{u}\widehat{z}'$ is an edge on the path $P$ and thus
    uncolored.  Since $\widehat{z}'\prec_{\TP}\widehat{u}$ and
    $C\setminus L(\TP(\widehat{u}))\ne\emptyset$, we also have
    $C\setminus L(\TP(\widehat{z}'))\ne\emptyset$.  As a consequence, we
    must have $C\cap L(\TP(\widehat{z}'))=\emptyset$, since otherwise
    $\widehat{u}\widehat{z}'$ would be colored by $C$.  Now
    $C\cap L(\TP(\widehat{z}))=C\cap L(\TP(\widehat{z}'))=\emptyset$ for
    two distinct children
    $\widehat{z},\widehat{z}'\in \child_{\TP}(\widehat{u})$ and
    $C\cap L(\TP(\widehat{u}))\ne\emptyset$ imply the existence of a third
    child
    $\widehat{z}''\in\child_{\TP}(\widehat{u})\setminus \{\widehat{z},
    \widehat{z}'\}$.

    Hence, we can consider the basic refinement step on
    $V'\coloneqq\{\widehat{z}, \widehat{z}'\}$, which introduces the set
    $W\coloneqq L(\TP(\widehat{z})) \cup L(\TP(\widehat{z}))$ into
    $\mathcal{H}(\TP)$.  Suppose $W$ is colored, i.e., there is set
    $D\in\mathcal{P}$ such that $D\cap W\ne \emptyset$ and
    $D\setminus W\ne \emptyset$.  Assume first that
    $L(\TP(\widehat{z}))\cap W\ne \emptyset$. Then
    $L(\TP(\widehat{z}))\subset W$ and $D\setminus W\ne \emptyset$ imply
    that $L(\TP(\widehat{z}))\setminus W\ne \emptyset$, and thus,
    $L(\TP(\widehat{z}))$ is colored by $D$; a contradiction to
    $\widehat{u}\widehat{z}$ being an uncolored edge.  Similar arguments
    rule out that $L(\TP(\widehat{z}'))\cap W\ne \emptyset$.  Therefore,
    such a set $D\in\mathcal{P}$ cannot exist and thus $W$ must be
    uncolored.  In summary, $\TP$ admits the uncolored basic
    refinement step on $V'$, contradicting the assumption that
    $\TP$ is a URS-tree.

    Thus neither case~(a) nor case~(b) is possible. The
    path $P$ in $\TP$ therefore does not contain a colored
    vertex $\widehat{v}\in V^0(\TP)$ such that $\lca_{\TP}(B)\prec_{\TP}
    \widehat{v}\preceq_{\TP} \widehat{u}$.
  \end{claim-proof}
  In summary, $\lca_{\TP}(A)$ and $\lca_{\TP}(B)$ are
  $\preceq_{\TP}$-incomparable and the path $P$ in $\TP$ does not
  contain a colored vertex $\widehat{v}\in V^0(T)$ such that
  $\lca_{\TP}(B)\prec_{\TP} \widehat{v}\preceq_{\TP} \widehat{u}=
  \lca_{\TP}(A\cup B)$.  By Thm.~\ref{thm:direction}(3), $(A,B)$ is
  ambiguous w.r.t.\ $\TP$ and $\mathcal{P}$.  Together with the fact that
  $\TP$ is a refinement of $T$ and thus compatible with $\mathcal{P}$ this
  implies that $(A,B)$ is r-ambiguous w.r.t.\ $T$ and $\mathcal{P}$.
\end{proof}

We summarize Prop.~\ref{prop:unambiguous-refinement} and
Lemma~\ref{lem:ambiguous-refinement} in the following characterization of
edges $(A,B)$ that are (un)ambiguously present or absent in refinements of
a given tree $T$ and valid choices of $H$:
\begin{theorem}
  \label{thm:direction-refinement}
  Let $T$ and $\mathcal{P}$ be r-compatible, $A,B\in\mathcal{P}$ be
  distinct, and $u\coloneqq \lca_{T}(A\cup B)$. Then the following
  statements hold:
  \begin{description}
  \item[(1)] $(A,B)$ is r-essential if and only if
    \begin{description}
    \item[(a)] the path $P_{u, \lca_{T}(B)}$ contains a colored edge, or
    \item[(b)] $u\ne\rho_T$ and the path $P_{u, \lca_{T}(A)}$ contains a
      colored edge $e$ with
      $\gamma(e)=\gamma(\parent_T(u) u)=C\in\mathcal{P}$.
    \end{description}
    \item[(2)] $(A,B)$ is r-forbidden if and only if
    \begin{description}
    \item[(c)] $\lca_{T}(A) \prec_T \lca_{T}(B)=u$ and $\gamma(uv)=B$ for
      the vertex $v\in \child_{T}(u)$ satisfying $\lca_{T}(A)\preceq_T v$.
    \end{description}
    \item[(3)] $(A,B)$ is r-ambiguous if and only if
    \begin{description}
      \item[(d')] the path $P_{u, \lca_{T}(B)}$ contains no colored edge,
        and 
      \item[(d'')] $u=\rho_T$ or the path $P_{u, \lca_{T}(A)}$ contains no
        colored edge $e$ with
        $\gamma(e)=\gamma(\parent_T(u) u)=C\in\mathcal{P}$, and
      \item[(d''')] $\lca_T(A)\not\prec\lca_T(B)$ or
        $\gamma(\lca_{T}(B)v)\ne B$ for the vertex
        $v\in \child_{T}(\lca_{T}(B))$ satisfying $\lca_{T}(A)\preceq_T v$.
    \end{description}
  \end{description}
\end{theorem}
\begin{proof}
  (1) Prop.~\ref{prop:unambiguous-refinement}(i) and~(iii) imply the
  \emph{if}-direction of statement~(1).  Assume, for contraposition, that
  the path $P_{u, \lca_{T}(B)}$ contains no colored edge and $u\ne\rho_T$
  or the path $P_{u, \lca_{T}(A)}$ contains no colored edge $e$ with
  $\gamma(e)=\gamma(\parent_T(u) u)=C\in\mathcal{P}$.  By
  Lemma~\ref{lem:ambiguous-refinement}, $(A,B)$ is r-forbidden or
  r-ambiguous, and thus not r-essential.

  (2) Prop.~\ref{prop:unambiguous-refinement}(ii) implies the
  \emph{if}-direction of statement~(2).  Suppose, for contraposition that
  $\lca_{T}(A) \not\prec_T \lca_{T}(B)=u$ or $\gamma(uv)\ne B$ for the
  vertex $v\in \child_{T}(u)$ satisfying $\lca_{T}(A)\preceq_T v$.  If the
  path $P_{u, \lca_{T}(B)}$ contains a colored edge or $u\ne\rho_T$ and the
  path $P_{u, \lca_{T}(A)}$ contains a colored edge $e$ with
  $\gamma(e)=\gamma(\parent_T(u) u)=C\in\mathcal{P}$, then, by
  statement~(1), $(A,B)$ is r-essential.  Otherwise,
  Lemma~\ref{lem:ambiguous-refinement} yields that $(A,B)$ is r-ambiguous.
  In any case, therefore, $(A,B)$ is not r-forbidden.

  (3) Lemma~\ref{lem:ambiguous-refinement} implies the \emph{if}-direction
  of statement~(3).  Assume, for contraposition, that one of the three
  conditions (d'), (d''), or (d''') in statement~(3) is not satisfied.  If
  the path $P_{u, \lca_{T}(B)}$ contains a colored edge, or $u\ne\rho_T$
  and the path $P_{u, \lca_{T}(A)}$ contains a colored edge $e$ with
  $\gamma(e)=\gamma(\parent_T(u) u)=C\in\mathcal{P}$, then $(A,B)$ is
  r-essential by statement~(1).  If neither is the case, we must have
  $\lca_T(A)\prec\lca_T(B)$ and $\gamma(\lca_{T}(B)v)= B$ for the vertex
  $v\in \child_{T}(\lca_{T}(B))$ satisfying $\lca_{T}(A)\preceq_T
  v$. Hence, we infer from statement~(2) that $(A,B)$ is r-forbidden.
  Thus $(A,B)$ cannot be r-ambiguous in any of the three cases.
\end{proof}

As alluded to in Section~\ref{sec:overview}, considering all refinements of
a tree $T$ (that is already compatible with a partition $\mathcal{P}$)
rather than $T$ alone introduces ambiguities in the classification of pairs
$(A,B)$ as (un)ambiguously present or absent in the Fitch graph.  This
observation is reflected in a comparison of Thms.~\ref{thm:direction}
and~\ref{thm:direction-refinement}: Condition~(a) (or (b)) in
Thm.~\ref{thm:direction-refinement} implies the existence of a colored
vertex $v\in V^0(T)$ such that $\lca(B)\prec v\preceq \lca(A\cup B)$ and
thus that $(A,B)$ is essential (cf.\ Thm.~\ref{thm:direction}(1)).
Similarly, condition~(c) in Thm.~\ref{thm:direction-refinement} implies
that $\lca_{T}(A) \prec_T \lca_{T}(B)$ and thus that $(A,B)$ is forbidden
(cf.\ Thm.~\ref{thm:direction}(2)).  The converses, however, are not true.

\begin{corollary}
  \label{cor:compl-refinement}
  Let $T$ and $\mathcal{P}$ be r-compatible. Then it can be decided in
  constant time after an $O(|L|)$ preprocessing step whether for
  distinct $A,B\in\mathcal{P}$ the pair $(A,B)$ is r-essential,
  r-forbidden, or r-ambiguous for $(T,\mathcal{P})$.
\end{corollary}

\begin{proof}
  By \cite[Thm.~7.5]{Hellmuth:21q}, the edge coloring $\gamma$ can be
  computed in $O(|L|)$ time for the r-compatible pair $(T,\mathcal{P})$.
  This step includes the computation of $\lca_{T}(A)$ for all
  $A\in\mathcal{P}$ and the construction an LCA data structure such as that
  of \citet{Bender:05} that enables constant-time queries for the last
  common ancestor for any pair $v,v'\in V(T)$.  In particular,
  $\lca_{T}(A\cup B)=\lca_{T}(\lca_{T}(A), \lca_{T}(B))$ for two sets
  $A,B\in\mathcal{P}$ is obtained in constant time.

  In order to evaluate the conditions in
  Thm.~\ref{thm:direction-refinement}, we need to access the vertex
  $w\in\child_{T}(u)$ satisfying $v\preceq_{T} w$ for two given vertices
  $u,v\in V(T)$ with $v\prec_{T} u$. To facilitate such queries we
  first determine $\depth(v)$ for each $v\in V(T)$, i.e., the number of
  edges on the path from the root to $v$. The values of $\depth(v)$ can
  be pre-computed by top-down traversal of $T$ in $O(|L|)$ time.  The
  \emph{Level Ancestor (LA) Problem} asks for the ancestor $\LA(v,d)$ of a
  given vertex $v$ that has depth $d$, and has solutions with $O(|L|)$
  preprocessing and $O(1)$ query time \cite{Berkman:94,Bender:04}.  Hence,
  for $u,v\in V(T)$ with $v\prec_{T} u$, we can obtain the vertex
  $w\in\child_{T}(u)$ satisfying $v\preceq_{T} w$ as $\LA(v,\depth(u)+1)$
  in constant time.

  We continue by computing, for each $v\in V(T)$, the unique colored edge
  $uw$ (provided it exists) along the path $P_{\rho,v}$ from the root
  $\rho$ to $v$ such that there is no other edge $u'w'\in E(P_{\rho,v})$
  with $w'\prec_{T} w$.  We achieve this by filling a map \texttt{lce} (for
  ``lowest colored edge'') in an $O(|L|)$ top-down recursion on $T$. More
  precisely, we initialize $\texttt{lce}(\rho)=\emptyset$, and then set
  $\texttt{lce}(v)\leftarrow v$ if $\gamma(\parent_T(v)v)\ne\emptyset$ and
  $\texttt{lce}(v)\leftarrow \texttt{lce}(\parent_T(v))$ otherwise.  For
  two vertices $u,v\in V(T)$ with $v\prec u$, there is a colored edge $e$
  on the path $P_{u,v}$ iff $\texttt{lce}(v)\prec_{T} u$, which is
  equivalent to $\texttt{lce}(v)\ne u$ and $\lca_{T}(\texttt{lce}(v),u)=u$.
  We can therefore query in constant time, whether or not such a colored
  edge $e$ exists for $u$ and $v$.

  As a consequence, we can evaluate condition~(a) and (d') in
  Thm.~\ref{thm:direction-refinement} in constant time.  Suppose
  condition~(b) in Thm.~\ref{thm:direction-refinement} is satisfied for two
  sets $A,B\in\mathcal{P}$. That is, $u\coloneqq\lca_{T}(A\cup B)\ne \rho$
  and the path $P_{u, \lca_{T}(A)}$ contains a colored edge $e$ with
  $\gamma(e)=\gamma(\parent_T(u) u)=C$ for some $C\in\mathcal{P}$.  The
  condition $u\ne\rho$ is checked in constant time. Moreover,
  $\gamma(e)=\gamma(\parent_T(u) u)=C$ and $e\in E(P_{u, \lca_{T}(A)})$
  imply that the edge $uw$, with $w\in\child_{T}(u)$ and
  $\lca_{T}(A)\preceq_{T} w$, lies on the path connecting two elements in
  $C$, and thus $\gamma(uw)=C$. It therefore suffices to check
  $\gamma(uw)=\gamma(\parent_T(u) u)=C\in\mathcal{P}$ for
  condition~(b)/(d''). This can be achieved in constant time using the LCA
  and LA data structures.  Similarly, checking condition~(c)/(d''') in
  Thm.~\ref{thm:direction-refinement} requires only constant-time queries.

  In summary, following $O(|L|)$ a preprocessing step, the conditions of
  Thm.~\ref{thm:direction-refinement} can be evaluated in constant time for
  each of the $O(|\mathcal{P}|^2)$ pairs $(A,B)$.
\end{proof}
The total effort to determine for all distinct $A,B \in \mathcal{P}$
whether $(A,B)$ is r-essential, r-forbidden, or r-ambiguous for
$(T,\mathcal{P})$ is therefore $O(|L| + |\mathcal{P}|^2)$.

\section{Computational Results}
\label{sect:comp}

In practical applications, estimates $\hat T$ and $\hat{\mathcal{P}}$ of
the underlying ``true'' gene tree $T$ and the ``true'' partition
$\mathcal{P}$ of the genes into HGT-free subsets, resp., can be obtained by
comparing the DNA or amino acid sequences of the genes in $L$. For the gene
trees $\hat T$, this is achieved either by standard methods of molecular
phylogenetics, reviewed e.g.\ by \citet{Yang:12}, or methods based on
pairwise best matches, see e.g.\ \cite{Hellmuth:15a}.  Estimates of
$\mathcal{P}$ can be obtained using indirect methods that compare the
divergence time of a pair of genes with genome-wide expectations
\cite{Novichkov:04,Ravenhall:15}.  In contrast, no methods to approximate
the directed Fitch graph $\fitch(T,H)$ from sequence data have become
available so far. The mathematical results above show that the separating
set $H$, and thus also $\fitch(T,H)$, are already determined -- at least in
part -- by $T$ and $\mathcal{P}$.  Naturally, this begs the question how
accurately $T$ and $\mathcal{P}$ determine $H$ and $\fitch(T,H)$ in
realistic scenarios. This is quantified conveniently by the fraction of
essential, ambiguous, and forbidden edges in $T$, and the fraction of
essential, ambiguous, and forbidden gene pairs $(x,y)$ in
$\fitch(T,H)$. Since we are interested in the theoretical limits of the
approach, we consider idealized conditions in which all sources of noise
and biases present in real-life sequence data are excluded. We therefore
consider simulated data in which the gene tree $T$ and HGT-induced
partition $\mathcal{P}$ are known.

To this end, we generated gene family histories (GFHs) that cover a wide
range of horizontal gene transfer (HGT) rates using the simulation library
\texttt{AsymmeTree} \cite{Stadler:20a}.  In brief, each GFH is obtained as
follows: First, a planted species tree with a user-defined number of leaves
(here drawn at random and independently between 10 and 100) is simulated
and endowed with a time map such that all leaves have a distance of one
time unit from the root.  In a second step, a gene tree is simulated along
the species tree using a constant-rate birth-death process with
user-defined rates for duplication, loss, and HGT events.  For all HGT
events, the recipient branch in the species tree is chosen at random among
the simultaneously existing branches.  Finally, all branches leading to
loss events only are removed to obtain the gene tree $T$.  We note that all
simulated gene trees are binary.  We simulated 5000 GFHs for various
combinations of event rates (indicated as triples on the horizontal axes of
the plots).

\begin{figure}
  \begin{center}
    \includegraphics[width=0.85\textwidth]{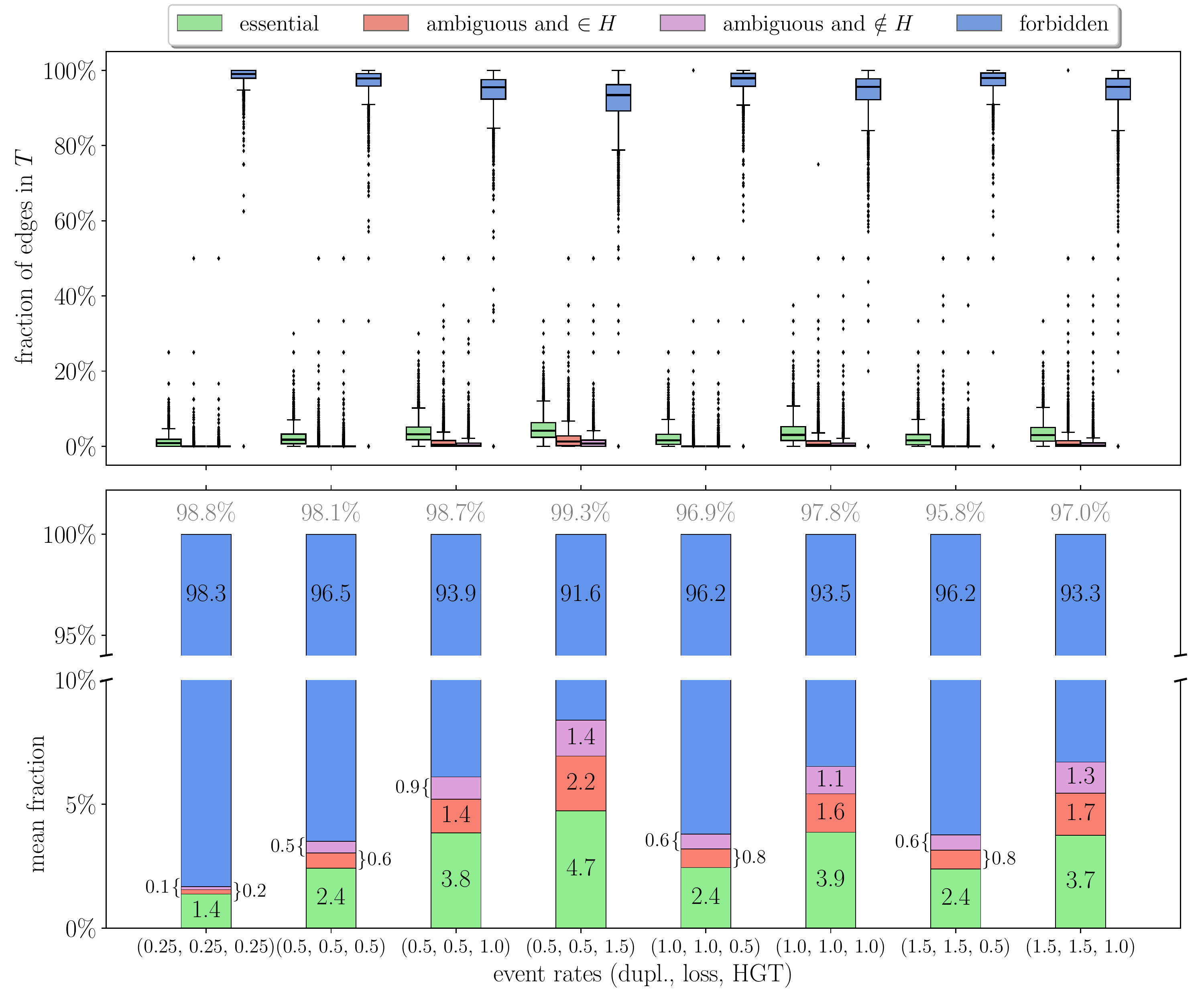}
  \end{center}
  \caption{Classification of edges in the gene trees $T$ based on the true
    undirected Fitch graph (represented by the partition $\mathcal{P}$).
    The tuples on the horizontal axis give the rates for duplication, loss,
    and HGT events.  Top panel: Fractions of four classes of edges:
    essential edges, ambiguous edges that are contained in $H$, ambiguous
    edges that are not contained in $H$, and forbidden edges.  Lower panel:
    Mean values of the fractions of the four classes.  The gray numbers are
    the proportions of scenarios (out of 5000 per rate combination) that
    were included in this analysis, i.e., the ones with $|L|>1$.}
  \label{fig:tree-edge-classification-bp}
\end{figure}

A GFH simulated in this manner contains the information of the types of
events for all vertices of $T$ as well as the separating set
$H\subseteq E(T)$ determined by the horizontal transfer events.  Using $T$
and $H$, we computed the partition $\mathcal{P}\coloneqq \partTH(T,H)$, and
classified all edges in $T$ as either essential, forbidden, or
ambiguous. Fig.~\ref{fig:tree-edge-classification-bp} shows the results in
terms of the fractions w.r.t.\ $|E(T)|$ (only GFHs with $|L|\ge2$ and thus
$|E(T)|>0$ were included, see gray fractions in the lower panel).  Not
surprisingly, the majority of edges lies on paths connecting leaves from
the same set in $\mathcal{P}$, i.e., they are forbidden.  Moreover, we
observe that there are on average more essential than ambiguous edges and
that more ambiguous edges are indeed contained in $H$ than not.

\begin{figure}
  \begin{center}
    \includegraphics[width=0.85\textwidth]{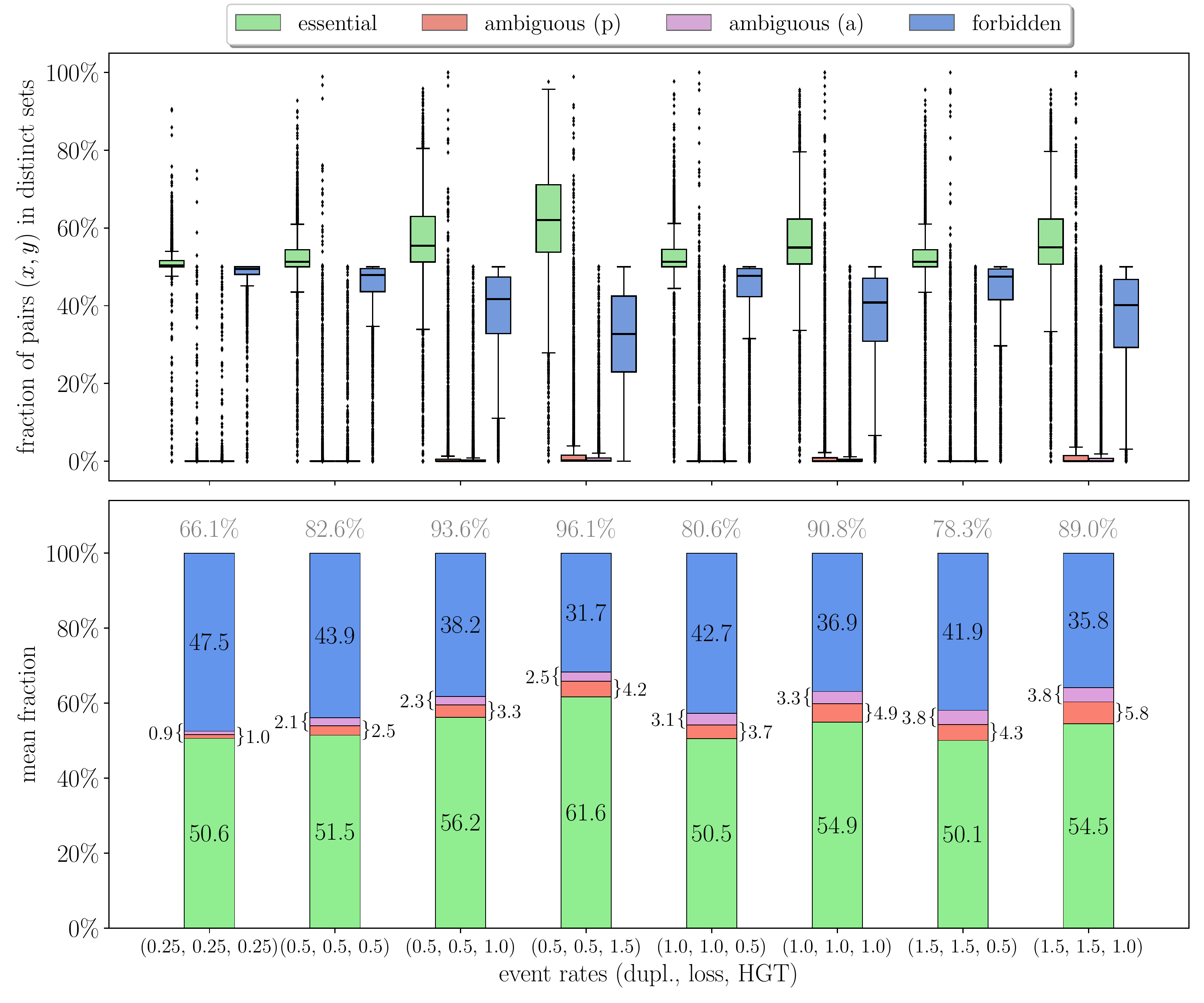}
  \end{center}
  \caption{Inference of edge orientation in the Fitch graphs of simulated
    scenarios on the basis of the true gene trees $T$ and the true
    undirected Fitch graph (represented by the partition $\mathcal{P}$).
    The tuples on the horizontal axis give the rates for duplication, loss,
    and HGT events.  Top panel: The ordered gene pairs are divided
    into four classes, whose relative abundance is displayed: essential
    edges, forbidden edges, ambiguous edges that are present (p) in true
    directed Fitch graph, and ambiguous edges that are absent from true
    directed Fitch graph. Lower panel: Mean values of the fractions of the 
    four classes.  The gray numbers are the proportions of scenarios (out of
    5000 per rate combination) that were included in the analysis, i.e.,
    the ones with $|\mathcal{P}|>1$.}
\label{fig:fitch-orientation-bp}
\end{figure}

We also classified all ordered pairs $(x,y)$ with $x\in A$ and $y\in B$ in
distinct sets $A,B\in\mathcal{P}$ as either \emph{essential},
\emph{forbidden}, or \emph{ambiguous} (in analogy to the respective
classification of $(A,B)$).  Since the true directed Fitch graph
$\fitch(T,H)$ is also known in the simulations, we can all determine for
all ambiguous pairs $(x,y)$ whether they are present or absent in
$\fitch(T,H)$.  Fig.~\ref{fig:fitch-orientation-bp} summarizes the results
of this edge classification.

The upper panel shows the distributions of the abundances of essential,
forbidden, and ambiguous edges as boxplots.  These relative values were
computed w.r.t.\ to the total number of pairs $(x,y)$ with $x$ and $y$ in
distinct sets of $\mathcal{P}$ in each scenario. Simulated GFHs without HGT
events (i.e., $|\mathcal{P}|=1$, $H=\emptyset$, and thus edge-less Fitch
graphs) are excluded from the quantitative analysis.  The fractions of GFHs
with $|\mathcal{P}|>1$ are indicated by the gray percentage values in the
lower panel of Fig.~\ref{fig:fitch-orientation-bp}.  The lower panel also
contains the mean proportions of essential, forbidden, and ambiguous edges.
To our surprise, $T$ and $\mathcal{P}$ unambiguously determine the presence
or absence of an edge in the Fitch graph for 90-98\% of the gene pairs
$(x,y)$, depending on the rates of events.

The accuracy of gene trees is inherently limited due to the limited number
of characters; collapsing poorly supported edges then results in minors of
the true, fully resolved gene tree.  Various sources of bias, furthermore,
may result in incorrectly inferred topologies even with full bootstrap
support, see e.g.\ \cite{Hahn:07,Som:15} and the references
therein. Several alternative approaches avoid the explicit reconstruction
of a gene trees and instead directly leverage comparisons of similarities
or distances to infer homology relations such as best matches and orthology
\cite{Setubal:18a,Altenhoff:19}. Minors of gene trees can be obtained as
``by-products'' of orthology \cite{Boecker:98,Hellmuth:13a} or the best
match relation \cite{Geiss:19a}. Usually, these are not fully resolved,
i.e., they can be obtained from the underlying true tree $T^*$ by a series
of inner edge contractions.  These trees contain partial but robust
information about $T^*$.  We consider here three distinct minors of $T^*$
that can be obtained in this manner: A unique discriminating cotree is
associated with orthology relations \cite{Hellmuth:13a}. Best-match
relations uniquely determine the least-resolved tree (LRT), see
\cite{Geiss:19a}, and the binary-resolvable tree (BRT). A BRT exists
whenever the underlying true gene tree $T^*$ was a binary tree
\cite{Schaller:20p}, which is the case in our simulations. In this setting,
our goal now is to classify pairs $(x,y)$ of genes as r-essential,
r-forbidden, or r-ambiguous.  Again we consider idealized conditions, i.e.,
we start from the true orthology and best match relations, which can be
extracted directly from the simulated GFHs.

\begin{figure}
  \begin{center}
    \includegraphics[width=0.85\textwidth]{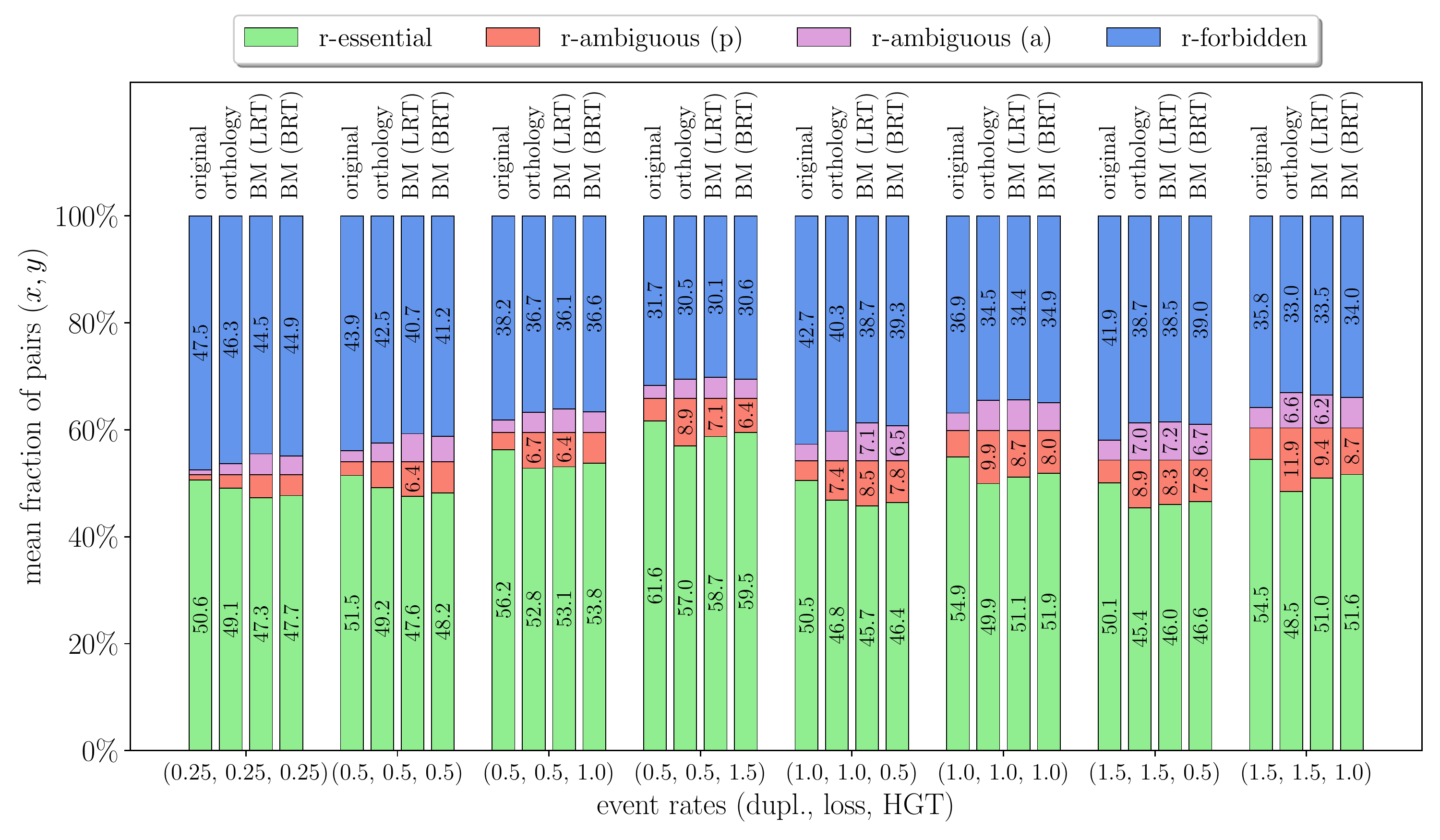}
  \end{center}
  \caption{Inference of edge orientation in the Fitch graphs taking into
    account incomplete resolution of the trees.  The same simulated
    scenarios as in Fig.~\ref{fig:fitch-orientation-bp} were included, and
    four trees were derived from them: the original gene tree (as in
    Fig.~\ref{fig:fitch-orientation-bp}), the discriminating cotree of the
    orthology relation, as well as two trees obtained from best matches
    (BM). The plot shows mean values of the fractions of the four classes:
    r-essential edges, r-ambiguous edges that correspond to present (p)
    edges in the true directed Fitch graph, r-ambiguous edges that
    correspond to absent (a) edges, and r-forbidden edges.}
  \label{fig:refinement-bp}
\end{figure}

Fig.~\ref{fig:refinement-bp} shows that the level of ambiguity remains
surprisingly small when these minors of $T$ obtained from orthology and
best matches are considered instead of the original, fully resolved tree.
On average, the vast majority of pairs are still classified as r-essential
and r-forbidden.  We note that, for binary trees, essential and r-essential
are equivalent. The same holds for forbidden and ambiguous explaining the
identical values in Fig.~\ref{fig:fitch-orientation-bp} (lower panel) and
Fig.~\ref{fig:refinement-bp} (original).

\section{Concluding Remarks}

Given a partition $\mathcal{P}$ of $L$ and a tree $T$ with leaf set $L$
that are compatible or at least r-compatible, we have obtained a complete
characterization of the essential, forbidden, and ambiguous pairs $(A,B)$
of distinct sets $A,B\in\mathcal{P}$. Furthermore, we have shown that this
classification can be computed in $O(|L|+|\mathcal{P}|^2)$ time for given
$T$ and $\mathcal{P}$. In biological terms, our result answers the question
to what extent the direction of horizontal gene transfers between
transfer-free subsets of genes (i.e., the sets of $\mathcal{P}$) are
already determined by a (not necessarily fully resolved) gene tree $T$: If
$(A,B)$ is essential, then for every $a\in A$ and $b\in B$, there is an HGT
event between $\lca(a,b)$ and $b$; if $(A,B)$ is forbidden, no such
HGT event can have taken place. In the ambiguous case, there are
evolutionary scenarios of both types. This classification is of practical
interest because the partition $\mathcal{P}$ of the gene set $L$ into
HGT-free subsets, i.e., the undirected (symmetrized) Fitch graph, can be
inferred from data \cite{Schaller:21f}, but so far no method has become
available to directly obtain the (directed) Fitch graph from
sequence or distance data. The mathematical results reported here thus
provide a means of locating HGT events on the gene tree and at the same time
to identify the ambiguities inherent in such a reconstruction.

From simulations of GFHs, we found that ambiguous edges are surprisingly
rare in Fitch graphs. As expected for randomly generated HGT events,
roughly half of the gene pairs are essential edges in the Fitch graph,
while the other half are forbidden pairs. Only a few percent of the pairs
are ambiguous. Regarding the edges in the gene tree, HGT can be ruled out
for most of them. Still, the majority of potential HGT edges are essential,
but the number of ambiguous edges is a sizeable fraction of $H^*$. The
results are qualitatively similar for the fully resolved gene tree and
several minors that can be inferred from orthology relations or best match
relations.  These numerical results indicate that HGT events can be
identified fairly accurately from $T$ and $\mathcal{P}$. It therefore
appears worthwhile to develop data analysis pipelines based on the results
obtained here.

Still, a subset of the HGT events remains ambiguous in general. This
begs the question whether there are additional sources of information
that could be employed to further reduce the ambiguities.  Several
avenues to improve the resolution are conceivable. First, one may ask
whether the LDT graph, which captures the experimental evidence more
directly than the undirected Fitch graph or its system of independent sets
$\mathcal{P}$, can be related directly to $T$. The LDT graph is always a
cograph and hence associated with a discriminating cotree $T^*$.  However,
this cotree $T^*$ is not necessarily displayed by the true gene tree
\cite{Schaller:21f}. Can one utilize the incompatibility of $T$ and $T^*$?

In practical applications, one usually also knows, for every $x\in L$, from
which species $\sigma(x)$ the gene was obtained. This coloring induces
additional constraints on the presence of HGT edges. For example, if two
distinct sets $A,B\in\mathcal{P}$ contain elements with same the vertex
color, i.e., if $\sigma(A)\cap\sigma(B)\ne\emptyset$, then there must be at
least two HGT edges on the path between $\lca(A)$ and $\lca(B)$. Can one
utilize this coloring information, which is also inherent in the LDT
graphs, in a systematic manner? The observation of \citet{Hellmuth:19} that 
Fitch graphs determine certain rooted triples that are necessarily displayed by 
species tree suggests that this should indeed be possible.

Finally, throughout this contribution we have assumed that
$\mathcal{P}$ is known but $T$ may be insufficiently resolved. What can be
said if our knowledge of $\mathcal{P}$ contains errors? It is always
possible to make $T$ and $\mathcal{P}$ compatible by refining $\mathcal{P}$
or coarsening $\mathcal{P}$, since every tree $T$ is compatible with both
the discrete and the indiscrete partition, i.e., with
$\{\{x\} \mid x\in L\}$ and $\{L\}$, respectively.  In the first case, we
have $H^*=E(T)$, in the second case $H^*=\emptyset$. What are the coarsest
refinements and the finest coarsenings of $\mathcal{P}$ that are compatible
with $T$? What can be said about the editing problem, if we assume that
both $T$ and $\mathcal{P}$ contain errors?

\section*{Acknowledgments}

This work was funded in part by the Deutsche Forschungsgemeinschaft
(proj.\ MI439/14-2).

\bibliography{preprint-directions}

\begin{thebibliography}{29}
\providecommand{\natexlab}[1]{#1}
\providecommand{\url}[1]{\texttt{#1}}
\expandafter\ifx\csname urlstyle\endcsname\relax
  \providecommand{\doi}[1]{doi: #1}\else
  \providecommand{\doi}{doi: \begingroup \urlstyle{rm}\Url}\fi

\bibitem[Altenhoff et~al.(2019)Altenhoff, Glover, and Dessimoz]{Altenhoff:19}
A.~M. Altenhoff, N.~M. Glover, and C.~Dessimoz.
\newblock Inferring orthology and paralogy.
\newblock In M.~Anisimova, editor, \emph{Evolutionary Genomics}, volume 1910 of
  \emph{Methods Mol. Biol.} Humana, New York, NY, 2019.
\newblock \doi{10.1007/978-1-4939-9074-0_5}.

\bibitem[Bender and Farach-Colton(2004)]{Bender:04}
M.~A. Bender and M.~Farach-Colton.
\newblock The level ancestor problem simplified.
\newblock \emph{Theor. Comput. Sci.}, 321\penalty0 (1):\penalty0 5--12, 2004.
\newblock \doi{10.1016/j.tcs.2003.05.002}.

\bibitem[Bender et~al.(2005)Bender, Farach-Colton, Pemmasani, Skiena, and
  Sumazin]{Bender:05}
M.~A. Bender, M.~Farach-Colton, G.~Pemmasani, S.~Skiena, and P.~Sumazin.
\newblock Lowest common ancestors in trees and directed acyclic graphs.
\newblock \emph{J. Algorithms}, 57\penalty0 (2):\penalty0 75--94, 2005.
\newblock \doi{10.1016/j.jalgor.2005.08.001}.

\bibitem[Berkman and Vishkin(1994)]{Berkman:94}
O.~Berkman and U.~Vishkin.
\newblock Finding level-ancestors in trees.
\newblock \emph{J. Comput. Syst. Sci.}, 48\penalty0 (2):\penalty0 214--230,
  1994.
\newblock \doi{10.1016/S0022-0000(05)80002-9}.

\bibitem[B{\"o}cker and Dress(1998)]{Boecker:98}
S.~B{\"o}cker and A.~W.~M. Dress.
\newblock Recovering symbolically dated, rooted trees from symbolic
  ultrametrics.
\newblock \emph{Adv. Math.}, 138:\penalty0 105--125, 1998.
\newblock \doi{10.1006/aima.1998.1743}.

\bibitem[Fitch(2000)]{Fitch:00}
W.~M. Fitch.
\newblock Homology: a personal view on some of the problems.
\newblock \emph{Trends Genet.}, 16:\penalty0 227--231, 2000.
\newblock \doi{10.1016/S0168-9525(00)02005-9}.

\bibitem[Gei{\ss} et~al.(2018)Gei{\ss}, Anders, Stadler, Wieseke, and
  Hellmuth]{Geiss:18a}
M.~Gei{\ss}, J.~Anders, P.~F. Stadler, N.~Wieseke, and M.~Hellmuth.
\newblock Reconstructing gene trees from {Fitch}'s xenology relation.
\newblock \emph{J. Math. Biol.}, 77:\penalty0 1459--1491, 2018.
\newblock \doi{10.1007/s00285-018-1260-8}.

\bibitem[Gei{\ss} et~al.(2019)Gei{\ss}, Ch{\'a}vez, Gonz{\'a}lez~Laffitte,
  L{\'o}pez~S{\'a}nchez, Stadler, Valdivia, Hellmuth, Hern{\'a}ndez~Rosales,
  and Stadler]{Geiss:19a}
M.~Gei{\ss}, E.~Ch{\'a}vez, M.~Gonz{\'a}lez~Laffitte, A.~L{\'o}pez~S{\'a}nchez,
  B.~M.~R. Stadler, D.~I. Valdivia, M.~Hellmuth, M.~Hern{\'a}ndez~Rosales, and
  P.~F. Stadler.
\newblock Best match graphs.
\newblock \emph{J. Math. Biol.}, 78:\penalty0 2015--2057, 2019.
\newblock \doi{10.1007/s00285-019-01332-9}.

\bibitem[Hahn(2007)]{Hahn:07}
M.~W. Hahn.
\newblock Bias in phylogenetic tree reconciliation methods: implications for
  vertebrate genome evolution.
\newblock \emph{Genome Biol.}, 8:\penalty0 R141, 2007.
\newblock \doi{10.1186/gb-2007-8-7-r141}.

\bibitem[Hellmuth(2017)]{Hellmuth:17}
M.~Hellmuth.
\newblock Biologically feasible gene trees, reconciliation maps and informative
  triples.
\newblock \emph{Alg. Mol. Biol.}, 12\penalty0 (23), 2017.
\newblock \doi{10.1186/s13015-017-0114-z}.

\bibitem[Hellmuth and Seemann(2019)]{Hellmuth:19}
M.~Hellmuth and C.~R. Seemann.
\newblock Alternative characterizations of {Fitch}'s xenology relation.
\newblock \emph{J. Math. Biol.}, 79:\penalty0 969--986, 2019.
\newblock \doi{10.1007/s00285-019-01384-x}.

\bibitem[Hellmuth and Wieseke(2016)]{HW:16book}
M.~Hellmuth and N.~Wieseke.
\newblock From sequence data incl.\ orthologs, paralogs, and xenologs to gene
  and species trees.
\newblock In P.~Pontarotti, editor, \emph{Evolutionary Biology -- Convergent
  Evolution, Evolution of Complex Traits, Concepts and Methods}, pages
  373--392, Cham, 2016. Springer.
\newblock \doi{10.1007/978-3-319-41324-2_21}.

\bibitem[Hellmuth et~al.(2013)Hellmuth, Hernandez-Rosales, Huber, Moulton,
  Stadler, and Wieseke]{Hellmuth:13a}
M.~Hellmuth, M.~Hernandez-Rosales, K.~T. Huber, V.~Moulton, P.~F. Stadler, and
  N.~Wieseke.
\newblock Orthology relations, symbolic ultrametrics, and cographs.
\newblock \emph{J. Math. Biol.}, 66:\penalty0 399--420, 2013.
\newblock \doi{10.1007/s00285-012-0525-x}.

\bibitem[Hellmuth et~al.(2015)Hellmuth, Wieseke, Lechner, Lenhof, Middendorf,
  and Stadler]{Hellmuth:15a}
M.~Hellmuth, N.~Wieseke, M.~Lechner, H.-P. Lenhof, M.~Middendorf, and P.~F.
  Stadler.
\newblock Phylogenetics from paralogs.
\newblock \emph{Proc. Natl. Acad. Sci. USA}, 112:\penalty0 2058--2063, 2015.
\newblock \doi{10.1073/pnas.1412770112}.

\bibitem[Hellmuth et~al.(2018)Hellmuth, Long, Gei{\ss}, and
  Stadler]{Hellmuth:18a}
M.~Hellmuth, Y.~Long, M.~Gei{\ss}, and P.~F. Stadler.
\newblock A short note on undirected {Fitch} graphs.
\newblock \emph{Art Discr. Appl. Math.}, 1:\penalty0 P1.08, 2018.
\newblock \doi{10.26493/2590-9770.1245.98c}.

\bibitem[Hellmuth et~al.(2021{\natexlab{a}})Hellmuth, Michel, N{\o}jgaard,
  Schaller, and Stadler]{Hellmuth:21w}
M.~Hellmuth, M.~Michel, N.~N. N{\o}jgaard, D.~Schaller, and P.~F. Stadler.
\newblock Combining orthology and xenology data in a common phylogenetic tree.
\newblock In P.~F. Stadler, M.~E. M.~T. Walter, M.~Hernandez-Rosales, and M.~M.
  Brigido, editors, \emph{Advances in Bioinformatics and Computational
  Biology}, volume 13063 of \emph{Lect. Notes Comp. Sci.}, pages 53--64, Cham,
  CH, 2021{\natexlab{a}}. Springer Nature.
\newblock \doi{10.1007/978-3-030-91814-9_5}.

\bibitem[Hellmuth et~al.(2021{\natexlab{b}})Hellmuth, Schaller, and
  Stadler]{Hellmuth:21q}
M.~Hellmuth, D.~Schaller, and P.~F. Stadler.
\newblock Compatibility of partitions, hierarchies, and split systems.
\newblock \emph{Discr. Appl. Math.}, 2021{\natexlab{b}}.
\newblock to appear; arXiv 2104.14146.

\bibitem[Jones et~al.(2017)Jones, Lafond, and Scornavacca]{Jones:17}
M.~Jones, M.~Lafond, and C.~Scornavacca.
\newblock Consistency of orthology and paralogy constraints in the presence of
  gene transfers.
\newblock 2017.
\newblock arXiv 1705.01240.

\bibitem[Lafond and Hellmuth(2020)]{lafond2020reconstruction}
M.~Lafond and M.~Hellmuth.
\newblock Reconstruction of time-consistent species trees.
\newblock \emph{Alg. Mol. Biol.}, 15\penalty0 (1):\penalty0 1--27, 2020.
\newblock \doi{10.1186/s13015-020-00175-0}.

\bibitem[N{\o}jgaard et~al.(2018)N{\o}jgaard, Gei{\ss}, Merkle, Stadler,
  Wieseke, and Hellmuth]{nojgaard2018time}
N.~N{\o}jgaard, M.~Gei{\ss}, D.~Merkle, P.~F. Stadler, N.~Wieseke, and
  M.~Hellmuth.
\newblock Time-consistent reconciliation maps and forbidden time travel.
\newblock \emph{Alg. Mol. Biol.}, 13\penalty0 (1):\penalty0 1--17, 2018.
\newblock \doi{10.1186/s13015-018-0121-8}.

\bibitem[Novichkov et~al.(2004)Novichkov, Omelchenko, Gelfand~Mikhail, Mironov,
  Wolf, and Koonin]{Novichkov:04}
P.~S. Novichkov, M.~V. Omelchenko, S.~Gelfand~Mikhail, A.~A. Mironov, Y.~I.
  Wolf, and E.~V. Koonin.
\newblock Genome-wide molecular clock and horizontal gene transfer in bacterial
  evolution.
\newblock \emph{J. Bacteriol.}, 186:\penalty0 6575--6585, 2004.
\newblock \doi{10.1128/JB.186.19.6575-6585.2004}.

\bibitem[Ravenhall et~al.(2015)Ravenhall, {\v{S}}kunca, Lassalle, and
  Dessimoz]{Ravenhall:15}
M.~Ravenhall, N.~{\v{S}}kunca, F.~Lassalle, and C.~Dessimoz.
\newblock Inferring horizontal gene transfer.
\newblock \emph{PLoS Comp. Biol.}, 11:\penalty0 e1004095, 2015.
\newblock \doi{10.1371/journal.pcbi.1004095}.

\bibitem[Schaller et~al.(2021{\natexlab{a}})Schaller, Gei{\ss}, Hellmuth, and
  Stadler]{Schaller:20p}
D.~Schaller, M.~Gei{\ss}, M.~Hellmuth, and P.~F. Stadler.
\newblock Best match graphs with binary trees.
\newblock In C.~Mart{\'\i}n-Vide, M.~A. Vega-Rodr{\'\i}guez, and T.~Wheeler,
  editors, \emph{Algorithms for Computational Biology, 8th AlCoB}, volume 12715
  of \emph{Lect. Notes Comp. Sci.}, pages 82--93, 2021{\natexlab{a}}.
\newblock \doi{10.1007/978-3-030-74432-8\_6}.

\bibitem[Schaller et~al.(2021{\natexlab{b}})Schaller, Lafond, Stadler,
  Wiesecke, and Hellmuth]{Schaller:21f}
D.~Schaller, M.~Lafond, P.~F. Stadler, N.~Wiesecke, and M.~Hellmuth.
\newblock Indirect identification of horizontal gene transfer.
\newblock \emph{J. Math. Biol.}, 83:\penalty0 10, 2021{\natexlab{b}}.
\newblock \doi{10.1007/s00285-021-01631-0}.
\newblock arXiv 2012.08897.

\bibitem[Semple and Steel(2003)]{Semple:03}
C.~Semple and M.~Steel.
\newblock \emph{Phylogenetics}.
\newblock Oxford University Press, Oxford UK, 2003.

\bibitem[Setubal and Stadler(2018)]{Setubal:18a}
J.~C. Setubal and P.~F. Stadler.
\newblock Gene phylogenies and orthologous groups.
\newblock In J.~C. Setubal, P.~F. Stadler, and J.~Stoye, editors,
  \emph{Comparative Genomics}, volume 1704, pages 1--28. Springer, Heidelberg,
  2018.
\newblock \doi{10.1007/978-1-4939-7463-4_1}.

\bibitem[Som(2015)]{Som:15}
A.~Som.
\newblock Causes, consequences and solutions of phylogenetic incongruence.
\newblock \emph{Briefings Bioinf.}, 16:\penalty0 536--548, 2015.
\newblock \doi{10.1093/bib/bbu015}.

\bibitem[Stadler et~al.(2020)Stadler, Gei{\ss}, Schaller, L{\'o}pez,
  Gonzalez~Laffitte, Valdivia, Hellmuth, and
  Hern{\'a}ndez~Rosales]{Stadler:20a}
P.~F. Stadler, M.~Gei{\ss}, D.~Schaller, A.~L{\'o}pez, M.~Gonzalez~Laffitte,
  D.~Valdivia, M.~Hellmuth, and M.~Hern{\'a}ndez~Rosales.
\newblock From pairs of most similar sequences to phylogenetic best matches.
\newblock \emph{Alg. Mol. Biol.}, 15:\penalty0 5, 2020.
\newblock \doi{10.1186/s13015-020-00165-2}.

\bibitem[Yang and Rannala(2012)]{Yang:12}
Z.~Yang and B.~Rannala.
\newblock Molecular phylogenetics: principles and practice.
\newblock \emph{Nature Rev. Genetics}, 13:\penalty0 303--314, 2012.
\newblock \doi{10.1038/nrg3186}.

\end{thebibliography}

%\clearpage 

\begin{appendix}
\section{Additional Figures}

\begin{figure}[h]
  \begin{center}
    \includegraphics[width=0.85\textwidth]{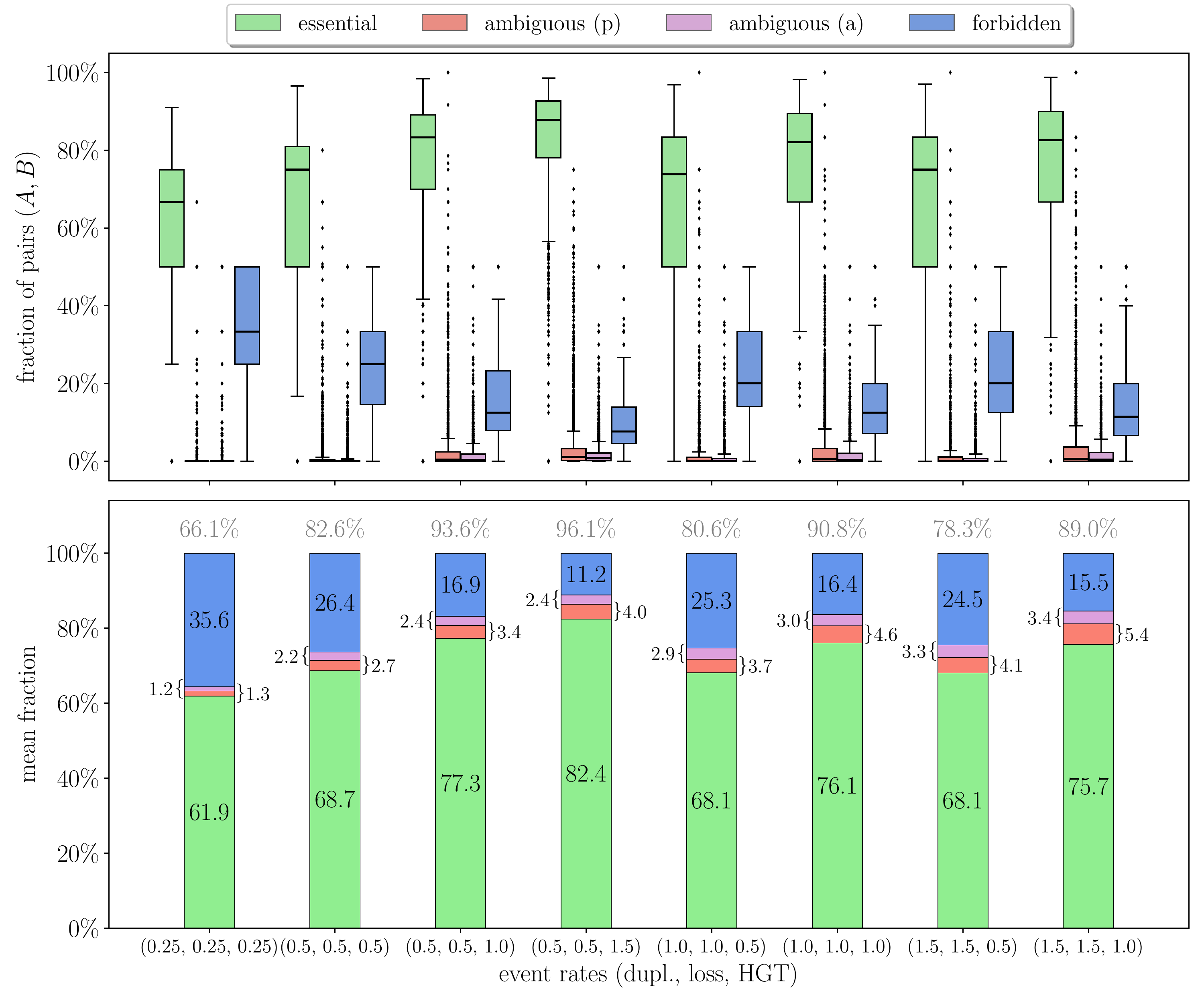}
  \end{center}
  \caption{Inference of edge orientation in the Fitch quotient graphs of
    simulated scenarios.  Top panel: Relative abundance of ordered pairs
    $(A,B)$ with $A,B\in\mathcal{P}$ that are essential, forbidden,
    ambiguous and present (p) in true Fitch quotient graph, and ambiguous
    and absent from true Fitch quotient graph. Lower panel: Mean values of
    the fractions of the four classes.  \textit{See also the caption of the
      analogous Fig.~\ref{fig:fitch-orientation-bp}.}}
  \label{fig:qfitch-orientation-bp}
\end{figure}

\begin{figure}[t]
  \begin{center}
    \includegraphics[width=0.85\textwidth]{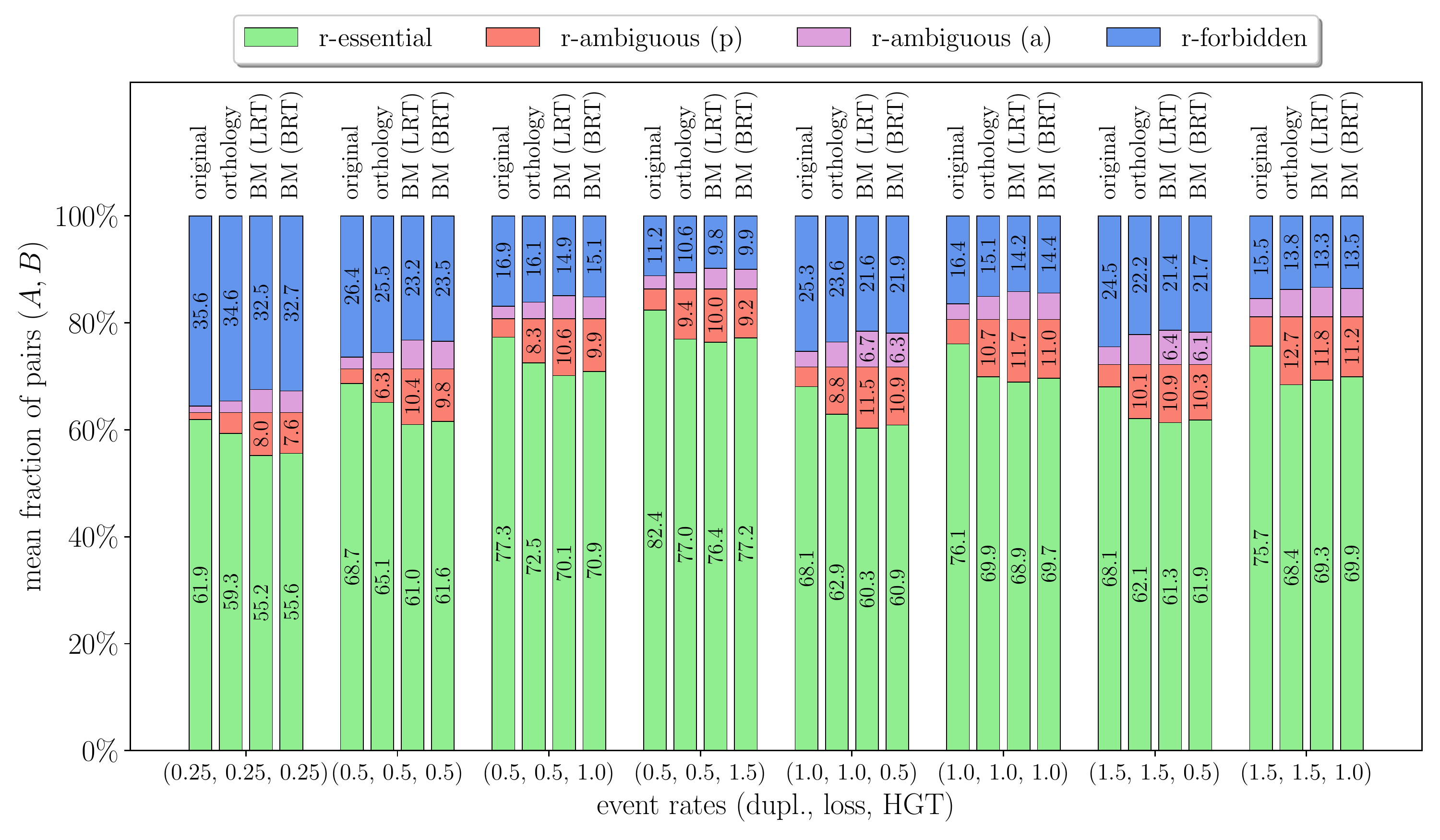}
  \end{center}
  \caption{Inference of edge orientation in the Fitch quotient graphs
    taking into account incomplete resolution of the trees.  The plot shows
    mean values of the fractions of pairs $(A,B)$ with $A,B\in\mathcal{P}$
    that are r-essential, r-ambiguous and present (p) in the true Fitch
    quotient graph, r-ambiguous and absent (a), and r-forbidden.
    \textit{See also the caption of the analogous
      Fig.~\ref{fig:refinement-bp}.}}
  \label{fig:qrefinement-bp}
\end{figure}

Figures~\ref{fig:fitch-orientation-bp} and~\ref{fig:refinement-bp}
show the fractions of (r-)essential, (r-)forbidden, and (r-)ambiguous
pairs of genes $(x,y)$ in Fitch graphs for simulated (r-)compatible trees
and partitions.  In this appendix, we display the same data for the pairs
of sets $(A,B)$ in the quotient Fitch graph. The relative abundances in
Fig.~\ref{fig:qfitch-orientation-bp} and Fig.~\ref{fig:qrefinement-bp},
resp., are qualitatively very similar, indicating that biases such as
imbalanced sizes of the sets $A$ and $B$ have little impact on the
overall conclusion that ambiguity is relatively rare plausible GFHs.

\end{appendix}
\end{document}